\documentclass[%
 reprint,groupedaddress,nofootinbib,nobibnotes,amsmath,amssymb,
aps,] {revtex4-1}

\usepackage{epstopdf}
\usepackage{eufrak}
\usepackage{amsmath,amssymb}
\usepackage{graphicx,epstopdf}
\usepackage{color}
\usepackage{subfigure}
\usepackage{enumitem}
\usepackage[dvipsnames]{xcolor}
\usepackage[utf8]{inputenc}
\usepackage{float}
\usepackage{mciteplus}
\usepackage{chngcntr}
\usepackage{booktabs}
\usepackage{bigints}
\usepackage{relsize}
\usepackage[english]{babel}
\renewcommand\harvardurl[1]{\textbf{URL}: \url{#1}} % optional
\usepackage{xurl} % allow line breaks anywhere in URL string
\usepackage{hyperref}

\usepackage{ragged2e}
\usepackage{array,tabularx}
\usepackage{makecell}
\newcolumntype{P}[1]{>{\centering\arraybackslash}p{#1}}
\begin{document}
	\title{Local and Non-local Fractional Porous Media Equations}
	
	\author{Fatemeh Gharari $^{[1]}$}
	\email{f.gharari@uma.ac.ir}
	\author{Karina Arias-Calluari $^{[2]}$}
	\author{Fernando Alonso-Marroquin $^{[2]}$}
	\author{Morteza. N. Najafi $^{[3]}$ }

\affiliation{$[1]$Department of Statistics, University of Mohaghegh Ardabili, Ardabil, Iran}
\affiliation{$[2]$School of Civil Engineering, The University of Sydney, Sydney NSW 2006}
\affiliation{$[3]$Department of Physics, University of Mohaghegh Ardabili, Ardabil, Iran}%

\begin{abstract}
Recently it was observed that the probability distribution of the price return in S\&P500 can  be  modeled by $q$-Gaussian distributions, where various phases (weak, strong super diffusion and normal diffusion) are separated by different fitting parameters (Phys Rev. E 99, 062313, 2019). Here we analyze the fractional extensions of  the porous media equation and show that all of them admit  solutions in terms of   generalized $q$-Gaussian functions. Three kinds of ``fractionalization'' are considered: \textit{local}, referring to the situation where the fractional derivatives for both space and time are local; \textit{non-local}, where both space and time fractional derivatives are non-local; and \textit{mixed}, where one derivative is local, and another is non-local.  Although, for the \textit{local} and \textit{non-local} cases we find $q$-Gaussian solutions , they differ in the number of free parameters. This makes differences to the quality of fitting to the real data. We test the results for the S\&P 500 price return and found that the local and non-local schemes fit the data better than the  classic  porous media equation.
\end{abstract}

	
	\maketitle

\section{Introduction}

The nonlinear diffusion equations have found vast applications in various fields in physics~\cite{Frank2}, neuroscience~\cite{Wedemann,Carrillo,Caceres}, psychology~\cite{Frank4}, economy and marketing~\cite{Michael,Borland2,Grandits,Ankudinova,Barles,Alonso}, biology and biophysics~\cite{Wedemann,Wedemann2,Chavanis,Frank2,Plastino}, and population dynamics~\cite{Okubo}. The examples of physical systems that are described by the nonlinear diffusion equation are the plasma systems~\cite{Escobedo,Takai}, surface physics~\cite{Spohn,Marsili}, astrophysics~\cite{Chavanis,Frank2}, the polymers~\cite{Ottinger,Ott}, fluids and particle beams~\cite{Plastino}, liquid surfaces~\cite{Bychuk}, nonlinear hydrodynamics~\cite{Barenblatt}, pattern formation~\cite{Parrondo} and laser physics~\cite{Kozyreff}, and most important, in financial analysis \cite{Alonso}. Many aspects of the nonlinear diffusion equation have been analyzed and found, like stationary~\cite{Wedemann2} and $H$-theorem~\cite{Shiino,Schwammle}, autocorrelations~\cite{Frank3}, path integral formulation~\cite{Wehner}, non-extensive maximum entropy approach~\cite{Borland,Martinez}, the distributed approximating functional method~\cite{Zhang}, the associated entropy~\cite{Schwammle2}, and anomalous diffusion~\cite{Tsallis}, for a good review see~\cite{Frank}. Due to the broadness of the problems involving anomalous diffusion, one needs to apply different kinds of theoretical approaches,such as the porous media equation (PME). The PME, as a subclass of the nonlinear anomalous diffusion equation, has been subjected to numerous and vast studies due to its possible applications for the porous media systems comprising of three essential equations: power-law equation of state, conservation of mass, and Darcy's law~\cite{Aronson}. Many analytical~\cite{Quiros,Gilding,Pamuk,Barbu,Peletier,Cordoba,Angenent,Ganji,Barbu2} and numerical~\cite{Bertsch,Duque,Macdonald,Ward} methods have been developed to study the properties of this model, which is not restricted to porous media systems, but also to the stock markets~\cite{Alonso,Bologna}. The fractional PMEs (FPMEs) has been studied in many papers~\cite{Bologna,Drazer,Compte,Tsallis2}, aiming to study anomalous diffusion in porous media and other problems related to PME, each of which with a particular (local or non-local) ``fractionalization'' scheme. Finding solutions for nonlinear anomalous diffusion equations is a challenge since, besides its difficulty to get exact analytical solutions, the principle of superposition is not applicable as in the linear case, so that the Fourier analysis can not be done.  Despite this huge interest and theoretical studies on the problem, very limited information is available concerning the possible solutions of these equations and their properties, especially the dependence of the solutions on the fractionalization parameters.\par In this paper, we aim to get to this issue by fractionalizing the PME with local and non-local fractional derivative operators. Intuitively a non-local operator is defined as the operator that needs the information in a finite interval upon its operation on a function, contrary to local operators that need only the information at one point in its close vicinity (see~\cite{ref1}, and Appendix~\ref{SEC:Katugampola-derivative}). We consider both local and non-local FPMEs focusing on three distinct cases: (LL) referring to the case where both time and space derivatives are local, (LN) or (NL) where one of them is local, and the other is non-local, and the (NN) referring to the case both derivatives of time and space are non-local. The main finding of the present paper is that the
   \textit{local} and \textit{non-local} cases    
  admit generalized $q$-Gaussian functions as their Green function solutions. The difference between them is the number and the form of the fitting parameters. As an application,  the local and non local generalized $q$-Gaussian distributions are used to describe the regimes observed during the time evolution of the probability density function (PDF) of the S\&P 500 index.\\
After addressing the $q$-Gaussian distribution function as a self-similar solution of the PME in the next section, we present the solution of the PME for the local fractional derivative (LL) in Sec.~\ref{SEC:local-local}. Sections~\ref{SEC:L-N} and~\ref{SEC:N-N}, contain the analysis of the PME with (LN) and (NN) fractionalization.   In Sec.~\ref{SEC:application}, we present an application of the generalized $q$-Gaussian distribution to describe the price return of S\&P 500 from the past 24 years, and we compare the results with previous solutions.

\renewcommand{\arraystretch}{4}
\begin{table*}[ht]
 \begin{tabular}{| p{0.5cm}| p{3cm}| p{13cm}| p{2cm}| } 
 \hline
 $N$ & Fractional Derivative & Definition & Ref\\
  \hline
1
& Katugampola & \scalebox{1.2} { ${\mathcal{D}} ^{\alpha}f(x)=\lim_{\epsilon \to 0}\dfrac{f^{(n)}(xe^{\epsilon x^{n-\alpha}})-f^{(n)}(x)}{\epsilon}\,\,\,\,\,( n <\alpha\leq n+1)$ }& \cite{Anderson2015} \\ 
 \hline
  2
& Riemann-Liouville & \scalebox{1.2} {  $ \begin{aligned}_{\,\,\,\,\,a}^{RL}{\mathcal{D}} ^{\alpha,x}f(x)&=\dfrac{1}{\Gamma(n-\alpha) }\left(\dfrac{d}{d x}\right)^n\bigintss_{a}^{x}\dfrac{f(\tau)}{(x-\tau)^{\alpha-n+1}}d\tau \,\,\,\,\,( n-1<\alpha\leq n)\\
 ^{RL}{\mathcal{D}}_{b}^{\alpha,x}f(x)&=\dfrac{1}{\Gamma(n-\alpha) }\left(-\dfrac{d}{d x}\right)^n\bigintss_{x}^{b}\dfrac{f(\tau)}{(\tau-x)^{\alpha-n+1}}d\tau\,\,\,\,\,( n-1<\alpha\leq n)\end{aligned}$}&\cite{Mainardi2006,Yang2019,Uchaikin2013}  \\ 

 \hline

3
& Caputo & \scalebox{1.2} {  $ \begin{aligned}_{\,\,a}^{C}{\mathcal{D}}   ^{\alpha,x}f(x)&=\dfrac{1}{\Gamma(n-\alpha) }\bigintss_{a}^{x}\dfrac{f^{(n)}(\tau)}{(x-\tau)^{\alpha-n+1}}d\tau\,\,\,\,\,( n-1<\alpha\leq n)\\
 ^{C}{\mathcal{D}}_{b}^{\alpha,x}f(x)&=\dfrac{(-1)^{n}}{\Gamma(n-\alpha) }\bigintss_{x}^{b}\dfrac{f^{n}(\tau)}{(\tau-x)^{\alpha-n+1}}d\tau\,\,\,\,\,( n-1<\alpha\leq n)\end{aligned}$}& \cite{Gorenflo1998,Yang2019,Uchaikin2013} \\ 
 \hline

 4
& Rietz & \scalebox{1.2} { $\begin{aligned}\left(\dfrac{d}{d \left|x\right|}\right)^{\alpha}f&=\dfrac{(^{RL}_{-\infty}\mathcal{D} ^{\alpha,x}+^{RL} \mathcal{D}_{\infty}^{\alpha,x})f(x)}{{2cos\left({\pi\alpha}/{2}\right)}} \\ %c_{\alpha}&=\dfrac{1}{2cos\left({\pi\alpha}/{2}\right)}\\ _{-\infty}\mathcal{D}_{x}^{\alpha}f(x)&=\dfrac{1}{\Gamma(n-\alpha)}\dfrac{\partial^{n}}{\partial x^{n}}\bigintss_{-\infty}^{x}\dfrac{f(\tau)}{(x-\tau)^{\alpha-n+1}}d{\tau}\\
%_{x}\mathcal{D}_{\infty}^{\alpha}f(x)&=\dfrac{(-1)^{n}}{\Gamma(n-\alpha)}\dfrac{\partial^{n}}{\partial x^{n}}\bigintss_{x}^{\infty}\dfrac{f(\tau)}{(\tau-x)^{\alpha-n+1}}d{\tau}
\end{aligned}$} 
& \cite{Gorenflo1998,Yang2019,Uchaikin2013,Gorenflo2002,Sun2018,Celik2012}\\ 
 \hline
\end{tabular}
\caption{Definitions of the most common fractional derivatives. }
\label{table:3}
\end{table*}
%\begin{table*}[ht]
 %\begin{tabular}{| p{3cm}| p{6cm}| p{6cm}| } 
 %\hline
 %& q=1 [independent] & q$\neq$1 (i.e, $Q\equiv2q-1 \neq 1$) [globally correlated]\\  
 %\hline
%$\begin{aligned}
%\sigma_{Q}&<\infty \\
%(\gamma&=2)\\
%\end{aligned}$
%& $\dfrac{\partial P}{\partial t}=D\dfrac{\partial^{2} P}{\partial x^{2}}$ (Bee) &  $\dfrac{\partial P}{\partial t}= D\dfrac{\partial^{2} P^{2-q}}{\partial x^{2}}$ (Dragonfly) \\ 
 % \hline
% $\begin{aligned}
%\sigma_{Q}& \rightarrow \infty \\
%(1<\gamma&<2)\\
%\end{aligned}$ &  $ \dfrac{\partial P}{\partial t}= D\dfrac{\partial^{\gamma} P}{\partial x^{\gamma}}$ (Spider)& $\dfrac{\partial^{\xi} P}{\partial t^{\xi}}= D\dfrac{\partial^{\gamma} P^{2-q}}{\partial x^{\gamma}}$ (Spider man) \\ 
% \hline
%\end{tabular}
%\caption{Summary of particular forms of the Partial Differential Equations (PDE). }
%\label{table:1}
%\end{table*}

%%%%%%%%%%%%%%%%%%%%%%%%%%%%%%%%
\section{A Fractional Generalization of PME }\label{SEC:OrdinaryPME}

The  PME  is one of the simplest examples of a nonlinear diffusion widely used to describe processes that involve fluid flows, gas flows, and heat transfer \cite{PME}.  The classical PME is:
\begin{equation}\label{ufo5}
\frac{\partial P}{\partial t}=D\frac{\partial^{2} P^{2-q}}{\partial x^{2}}.
\end{equation}
A solution of this partial differential equation (PDE) is the Barenblatt function for $q>1$ and $ t>0 $~\cite{ref9},
\begin{equation}\label{ufo6}
P(x,t)=\frac{1}{(Dt)^{\frac{1}{3-q}}} \left( C-\dfrac{1-q}{2(2-q)(3-q)}\frac{x^{2}}{(Dt)^{\frac{2}{3-q}}}\right) ^{\frac{1}{1-q}},
\end{equation}
%( I THINK YOU NEED TO BRING THE TABLE WITH THE DEFINITION OF THE DERIVATIVES  AND EXPLAIN DERIVATIVE WITH RESPECT TO A FUNCTION AFTER EQUATION 3.)
where $C$ is an integration  constant. The Eq.(\ref{ufo5}) has been generalized to analyze several physical situations that present anomalous diffusion \cite{Lenzi}. 
The present paper proposes a generalized form of the PME that admits a broader range of results:

\begin{equation}
\label{Eq:Matrix}
_{a}\mathcal{D} ^{\xi, t^{n}}P(x,t)=D  \left(^{C} \mathcal{D}_{b}^{\gamma, x^{\alpha}}P^{\nu}(x,t)\right).
\end{equation}

 The fractional derivative $\mathcal{D}$ of orders $\xi$ and $\gamma$ is a function of three variables: The limits ($a$ and $b$), the arguments ($t$ and $x$), and the degree order of the arguments ($n$ and $\alpha$). This last type of variable   allows us to have a derivative with respect to a function when $ n,\alpha \neq 1 $.   
\\
 A particular case of the Eq. (\ref{Eq:Matrix}), when $n=\alpha =1$ and $a=b=0,$ is the nonlinear fractional diffusion equation 
\begin{equation}
\begin{split}
&\dfrac{\partial^{\xi} }{\partial t^{\xi} }P(x,t)=D\dfrac{\partial^{\gamma}}{\partial x^{\gamma}}P^{\nu}(x,t).
\end{split}
\label{general eq.}
\end{equation}

%(KARINA: PLEASE FIX HERE THE CONCEPT OF GREEN FUNCTION THAT IS NOT CLEAR, THE GREEN FUNCTION IS THE SOLUTION OF THE DIFFERENTIAL EQUATI0N WITH INITIAL CONDITION P(x,t=0) =  delta(x) AND BOUNDARY CONDITION P(x=+ - INFINITE,t)
No general solution of Eq.(\ref{general eq.}) is known.  In the present paper, we aim to show
 that a particular solution to this PDE is the Green function. This function is  obtained from the boundary condition $P(x,t)=0$,  when $x\rightarrow \pm \infty$, and the initial condition $P(x,0)=\delta(x)$, where $\delta(x)$ is the Dirac delta function.  The Green functions can be expressed in terms of well-known distributions.  Some cases are the Gaussian, the Levy-Stable \cite{Lenzi}, and the $q$-Gaussian distributions.  We show that the Eq.(\ref{general eq.}) admits exact solutions that vary depending on the definition of the fractional derivatives applied. The definitions of the most commonly used fractional derivatives are contained in Table \ref{table:3}. The Eq.(\ref{general eq.}) allows space and time to scale differently, and as a consequence, different solutions can be obtained. 

\renewcommand{\arraystretch}{2.3}
\begin{table*}[htbp]
\centering
\begin{tabular}{ |P{0.15cm}|P{1.5cm}|P{5.05cm}|P{7.45cm}|P{1.9cm}|P{2cm}| } 
\hline
N & Equation & Definition & Green function & Fractional derivatives & Authors \\
\hline
1 & Diffusion (*)  & $\dfrac{\partial P(x,t)}{\partial t}=D\dfrac{\partial^{2} P (x,t)}{\partial x^{2}}$ & $\begin{aligned} P(x,t)&=\dfrac{1}{\sqrt{2Dt}}g \left( \dfrac{x}{\sqrt{2Dt}}\right),\\ g(x)&=\dfrac{1}{\sqrt{\pi}}e^{-x^{2}} \end{aligned}$ & Integer derivatives & Bachelier \cite{Bachelier},  A.Einstein \cite{Mandelbrot,Einstein}  \\ 
\hline
2 & Anomalous super-diffusion  & $\begin{aligned} \dfrac{\partial P(x,t)}{\partial t}=&D\dfrac{\partial^{\gamma} P (x,t)}{\partial x^{\gamma}},\\ 0<\gamma&<2 \end{aligned}$  & $\begin{aligned}P(x,t)&=\dfrac{1}{(D t)^{1/\gamma}}{L_{\gamma}\left( \dfrac{x}{(D t)^{1/\gamma}}\right), }\\ L_{\gamma}(x)&=\dfrac{1}{\pi} \int_{0}^{\infty}e^{-\alpha \left |k  \right | ^{\gamma}}cos(kx) dk \\
\end{aligned}$ & Riesz &  P.Levy \cite{Levy1954,Gorenflo1999}, W.Feller \cite{Feller1962,Gorenflo1999,Xu2019,Janakiraman2012}  \\ 
\hline
3 & Classical PME (*)  & $\begin{aligned}\dfrac{\partial P(x,t)}{\partial t}=&D\dfrac{\partial^{2} P (x,t)^{2-q}}{\partial x^{2} },\\
5/3<q&<3
\end{aligned}$ & $\begin{aligned}P(x,t)=\dfrac{1}{\sqrt{3-q}(D t)^{\frac{1}{3-q}}}&{g_{q}\left( \dfrac{x}{\sqrt{3-q}(D t)^{\frac{1}{3-q}}}\right), }\\g_{q}(x)=\dfrac{1}{C_{q}}e_{q}(-x^{2}) , \,\,\,\,\,\, e_{q}(x)&=[1+(1-q)x]^{\frac{1}{1-q}}, \\
Cq =\sqrt{\dfrac{\pi}{q-1}}& \dfrac{\Gamma((3-q)/(2(q-1)) ) }{\Gamma(1/(q-1))} \\
\end{aligned}$ & Integer derivatives & Barenblatt \cite{Barenblatt1952,Esteban1986} C.Tsallis \cite{Drazer,Tsallis2005,Plastino1995,Tsallis1996}     \\ 
\hline
4 & Space-FPME  & $\begin{aligned}
    \dfrac{\partial P(x,t)}{\partial t}&=D\dfrac{\partial^{\gamma} P (x,t)^{\nu}}{\partial x^{\gamma}},\\
    \nu&=\dfrac{2-\gamma}{1+\gamma}\\
\end{aligned}$ & $\begin{aligned}
P(x,t)&=\dfrac{1}{C_{q}(k_{1}t)^{(\gamma+1)/(\gamma^{2}-\gamma+1)}}{\left(\dfrac{z^{\gamma(\gamma+1)}}{(1+bz)^{1-\gamma^{2}}}\right)^{\frac{1}{1-2\gamma}}} \\
z&=\dfrac{x}{(\left|k_{1}\right|t)^{{(\gamma+1)}/{(\gamma^{2}-\gamma+1})}} \,\,\,\, b=cte,\\ \dfrac{1}{C_{q}}&=\left[k \dfrac{\Gamma(\beta)}{\Gamma(\alpha+1)} \right]^{\dfrac{1+\gamma}{1-2\gamma}},\\ \alpha &=\dfrac{(2-\gamma)\gamma}{(1-2\gamma)},  \,\,\,\, \beta=-\dfrac{\gamma^{2}-3\gamma+2}{1-2\gamma}  \\ \end{aligned}$  & Riemann-Liouville (RL) & C.Tsallis \cite{Tsallis,Bologna}  \\
 
\hline
5 & Time-FPME (*)  & $\dfrac{\partial^{\xi} P(x,t)}{\partial t^{\xi}}=D\dfrac{\partial^{2} P (x,t)^{2-q}}{\partial x^{2}}$ & $P(x,t)=\dfrac{1}{\left(\dfrac{D t^{\xi}}{\xi}\right)^{\frac{1}{3-q}}} g_{q}\left(\dfrac{x}{\left(\dfrac{D t^{\xi}}{\xi}\right)^{\frac{1}{3-q}}}\right)$ & Katugampola  & Eq.(\ref{Eq:Self_similar})  \\ 
\hline

6 & Time-Space-FPME (*)& $\begin{aligned}
\dfrac{\partial^{\xi}}{\partial t^{\xi}}P(x,t)&=D \dfrac{\partial^{\gamma}}{\partial x^{\gamma}}P^{2-q}(x,t),\\
0<\xi \leq 1, & \,\,1<\gamma \leq 2,\,\, 1<q<3 
\end{aligned}$ & $ \begin{aligned}P(x,t)&=\dfrac{1}{(B t)^{\frac{\xi}{1-q+\gamma}}}{g_{q}^{\gamma}\left( \dfrac{x}{(B t)^{\frac{\xi}{1-q+\gamma}}}\right)},\\
g_{q}^{\gamma}(x)=&\dfrac{1}{C_{q}^{\gamma}}e_{q} (-x^{\gamma}), \,\,\,\,\, e_{q}(x)=[1+(1-q)x]^{\frac{1}{1-q}},\\
C_{q}^{\gamma}&=\dfrac{2\Gamma \left(\frac{1}{q-1}-\frac{1}{\gamma}\right)\Gamma \left(1+\frac{1}{\gamma}\right)}{(q-1)^{1/\gamma}\Gamma\left( \frac{1}{q-1}\right)},\\
B&=-\left[\dfrac{\xi}{D \gamma (\gamma-1)(2-q)(1-q+\gamma)} \right]^{\frac{\gamma}{1-q+\gamma}}
\end{aligned}$ & Katugampola  & Eq.~(\ref{eq:p_x})  \\
\hline
7&  Time-Space-FPME    & $\begin{aligned}
&\dfrac{\partial^{\xi}}{\partial t^{\xi}}P(x,t)=D\dfrac{\partial^{\gamma}}{\partial x^{\gamma}}P^{\nu}(x,t),\\
& 0<\xi \leq 1,\,\,0<\gamma<\frac{1}{2},\,\, \nu>-1 
\end{aligned}$ & $P(x,t)=\dfrac{A}{t^{{1}/{\sigma}}}\left(\dfrac{x}{t^{{1}/{\sigma}}}\right)^{\alpha\gamma}\left(c_{1}+c_{2}\frac{x}{t^{{1}/{\sigma}}}\right)^{-\alpha(1-\gamma)}$& Katugampola (time) and RL (space) & Eq.~(\ref{ufo11}) \\
\hline
8& Particular case of the Generalized PME ($\star$)  & $ \begin{aligned}_{0}\mathcal{D}_{t}^{\xi,\frac{1}{t}}P(x,t)= D & \left(^{C}_{x}\mathcal{D}_{1}^{\gamma, x^{\alpha}}P^{\nu}(x,t)\right)\\ \xi,\gamma>0,&\\
\xi=1+\lambda&+1/\alpha, \\ \nu=-1-&\frac{2}{\lambda+1/\alpha},\\
\gamma=(1/\alpha-\lambda)& \left(1+\frac{1}{\lambda+1/\alpha}\right)\end{aligned}$ & $ \begin{aligned}P(x,t)&=\dfrac{1}{(B t)^{\frac{1}{\alpha}}}{g_{q}^{\alpha,\lambda}\left( \dfrac{x}{(B t)^{\frac{1}{\alpha}}}\right)},\\
g_{q}^{\alpha,\lambda}(x)=&\dfrac{1}{C_{q}^{\alpha,\lambda}}e_{q}^{\alpha,\lambda} (-x^{\alpha}) , \,\,\,\,\, e_{q}^{\lambda}(x)=[1+(1-q)x]^{\lambda},\\
C_{q}^{\alpha,\lambda}&=2 \lambda^{1/\alpha} \dfrac{ \Gamma \left(-\lambda-\dfrac{1}{\alpha} \right) \Gamma \left(1+\dfrac{1}{\alpha} \right)}{\Gamma(-\lambda)},\\
B&=-\dfrac{1}{\eta^{NL}_{q}(\alpha,\lambda,D)} \,\,\,\, Eq.(\ref{eq:nq})
\end{aligned}$ & RL (time) and Caputo (space) & Eq.~(\ref{ufo12}) \\
 \hline
\end{tabular}
\caption{Summary of  the FPMEs and their Green functions. Equations marked with (*) are used to fit the PDF  of the S\&P 500 index; while ($\star$) represents the best model. }

\label{table:1}
\end{table*}
In searching for the solutions of Eq.(\ref{general eq.}), we exploit the fact that they follow the self-similarity law.

\begin{equation}
\label{eq:scaling}
P(x,t)=\dfrac{1}{\phi(t)}f\left[\dfrac{x}{\phi(t)} \right].    
\end{equation}

The Eq.(\ref{eq:scaling}) has often been used to model the price return in stock markets. This price return obeys $\phi(t)\sim t^{H}$, being $H$ the characteristic exponent of the  PDF. Additionally, for stock markets, $f$ fits well to a $q$-Gaussian distribution \cite{Alonso}. Then, the equations presented in Table \ref{table:1} can be used to model the detrended price return if they obey the power law and if its solution is a $q$-Gaussian.\\

 A summary of these solutions are contained in Table \ref{table:1}.
 The first four equations of Table {\ref{table:1}} have been solved previously by applying integer or fractional derivatives \cite{Uchaikin2003,Bologna,Tsallis2005}. The last four equations have been solved in this manuscript after applying local and non-local derivatives. The local derivatives were used to solve equations N.$5$ and N.$6$ in Table \ref{table:1}. The solution of   equation  N.$6$ is the \textit{first generalized} $q$-Gaussian function. The fractional derivatives used to solve equations N.$7$ and N.$8$ were of non-local character. The equation N.$7$ in Table \ref{table:1} is proposed as an improvement of equation N.$4$.  The equation N.$7$ presents fractional derivatives on both arguments, time and space.
%(KARINA: WHICH SELF-SIMILAR LAW? I THINK ALL SATISFY A SELF SIMILAR LAW, ACCORDING TO TSALLIS PAPER. THE LAW THAT YOU WANT TO MENTION HERE IS THE ONE FOR STOCK MARKET. i WOULD SUGGEST TO WRITE DOWN BOTHS). 
After applying local derivatives on time and non-local derivatives on the space, the solution of equation N.$7$ is not a $q$-Gaussian, and does not satisfy the self-similar law, $P(x,t) \nsim t^{-H}F(xt^{-H})$. Then, a particular form of Eq.~(\ref{Eq:Matrix}) is proposed as equation N.$8$ in Table \ref{table:1}. The equation N.$8$ has been solved by applying non-local fractional definitions on time and space. The solution of Eq.~(\ref{Eq:Matrix}) is called \textit{second generalized} $q$-Gaussian function and it presents a self-similar form, $P(x,t) \sim t^{-H}F(xt^{-H})$. By replacing $\lambda=\frac{1}{1-q}$ in the \textit{second generalized} $q$-Gaussian,  the \textit{first generalized} $q$-Gaussian can be recovered. 
\\
In the present paper, we aim to show that both local and non-local FPME admit \textit{generalized} $q$-Gaussian solutions. For local derivation, the Katugamapola definition is applied; for non-local derivation, the Riemann-Liouville and Caputo derivatives are applied. For the non-local derivation, we consider a fractional derivative with respect to another
function, in the sense of Caputo derivative.
%This equation contains Eq.~\ref{ufo5} as a particular case when $\xi=1$ and $\gamma=2$. This equations seems to see the 
%No general solution
%of Eq.~\ref{general eq.} is yet known to the best of our knowledge.

%%%%%%%%%%%%%%%%%%%%%%%%%%%%%%%%
\section{PME with LL Fractional Operators}\label{SEC:local-local}
%Fractional calculus allows one to generalize derivatives integrals of an integer order to an arbitrary order. 
The generalized forms of PME are obtained by replacing the first time derivative or second space derivative by  fractional orders derivatives in the classical PME. 
These generalized PMEs may model more efficiently certain real-world phenomena, especially when the dynamics are affected by constraints inherent to the system. Typically, fractional derivatives are defined with an integral representation. Consequently, they are non-local in character. There exists several definitions for fractional derivatives and fractional integrals like the Riemann-Liouville, Caputo, Hadamard, Riesz, Grünwald-Letnikov. However, some usual properties of these fractional derivatives are different from ordinary derivatives, such as the Leibniz rule, the chain rule, and the semigroup property.
Consequently, these fractional derivatives can not be applied for local scaling or differentiability properties. For
further details, we refer the reader to~\cite{ref21,ref23} and Appendix \ref{SEC:Definition-Properties}.
\subsection{  Local fractional Operators}
The concept of local fractional derivatives keep some of the properties of ordinary derivatives.  Nevertheless they lose the memory condition of fractional order derivatives~\cite{ref22}.  There exist several definitions for local fractional derivatives like the Kolwankar, Chen, Conformable, Katugampola, see~\cite{ref1} for details.  Recently, these local derivatives have been used to model phenomena of turbulence ~\cite{ref10}, and anomalous diffusion~\cite{ref11}.

Local definitions for fractional derivatives  were applied by Lenzi et al.~\cite{ref4}, his work contains some classes of solutions of a general fractional nonlinear diffusion equation with some observations. A similar study was made by Assis et al.~\cite{ref5}.  More information about the use of local fractional derivatives can be found in~\cite{ref6}, and~\cite{ref7}. These examples are related to diffusion equations with nonlinear terms and fractional time derivatives which are quite few \cite{ref2,ref3}. \par 

One popular type of local fractional derivatives is the ``conformable operator"~\cite{ref12}. The ``conformable operator" has been used in a wide range of applications. Some applications of the ``conformable operator" are in Newtonian mechanics~\cite{ref13}, diffusion equation~\cite{ref14}, and  nonlinear diffusion equation~\cite{ref15}. However, this local operator cannot be applied with zero as an order of the derivative.  Recently, the Katugampola~\cite{ref17} operator has been used as a limit based fractional derivative that allows zero as a possible order of the derivative. The Katugampola operator maintains many of the familiar properties of standard derivatives such as the product, quotient, and chain rules. 
Throughout this section, we consider the Katugampola derivative (Katugampola operator), to solve the generalized PME. By applying this local derivative, our solution will be a generalized $q$-Gaussian  distribution.
 Information about the Katugampola's definition and its properties can be found in Table \ref{table:3}, Appendix~\ref{SEC:Katugampola-derivative} and Table \ref{table:4}.

\begin{figure*}[!htbp]
\centering 
\includegraphics[scale=0.40,trim=0cm 0cm 0cm 15cm]{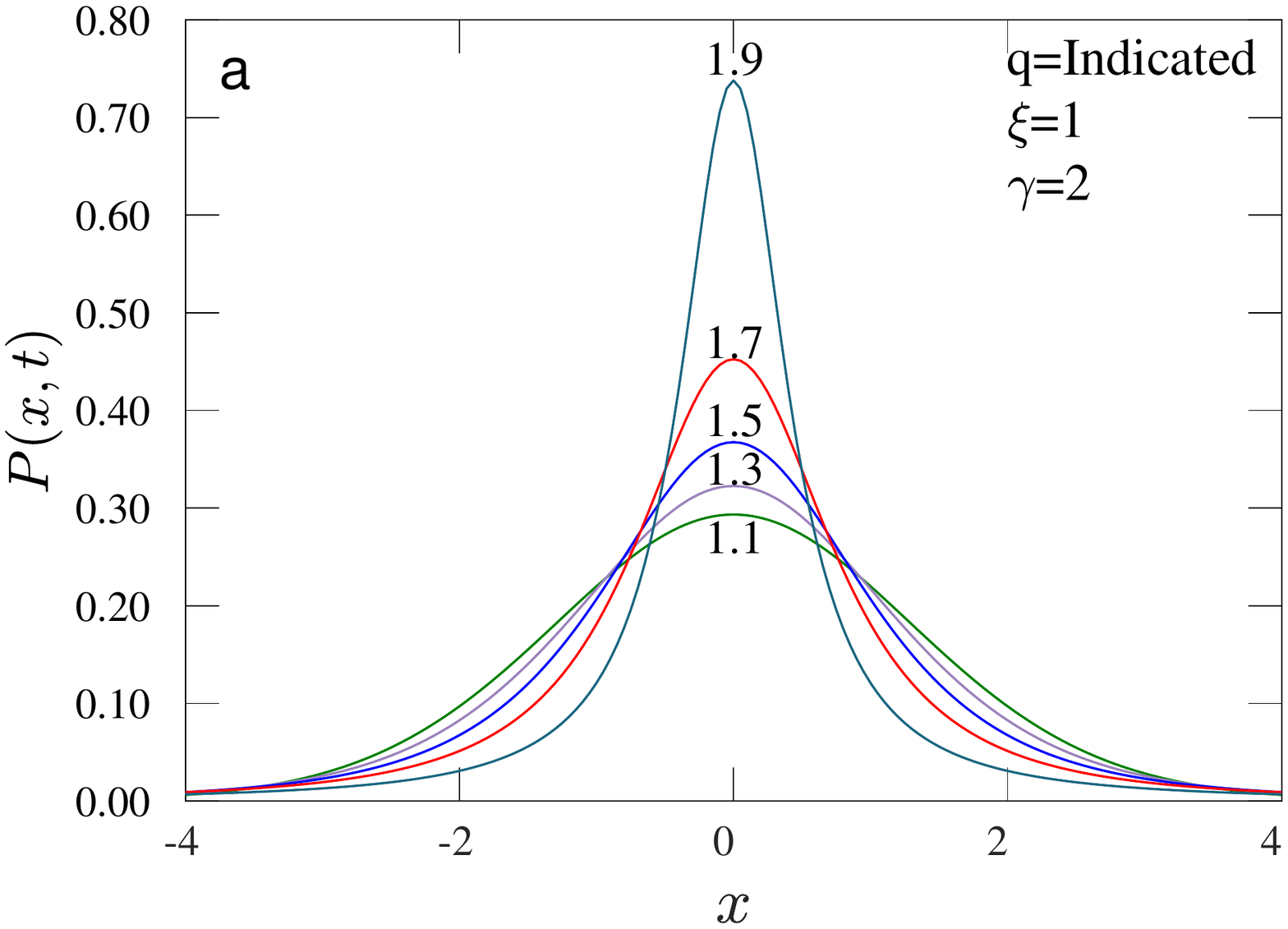}
\includegraphics[scale=0.40,trim=0cm 0cm 0cm 15cm]{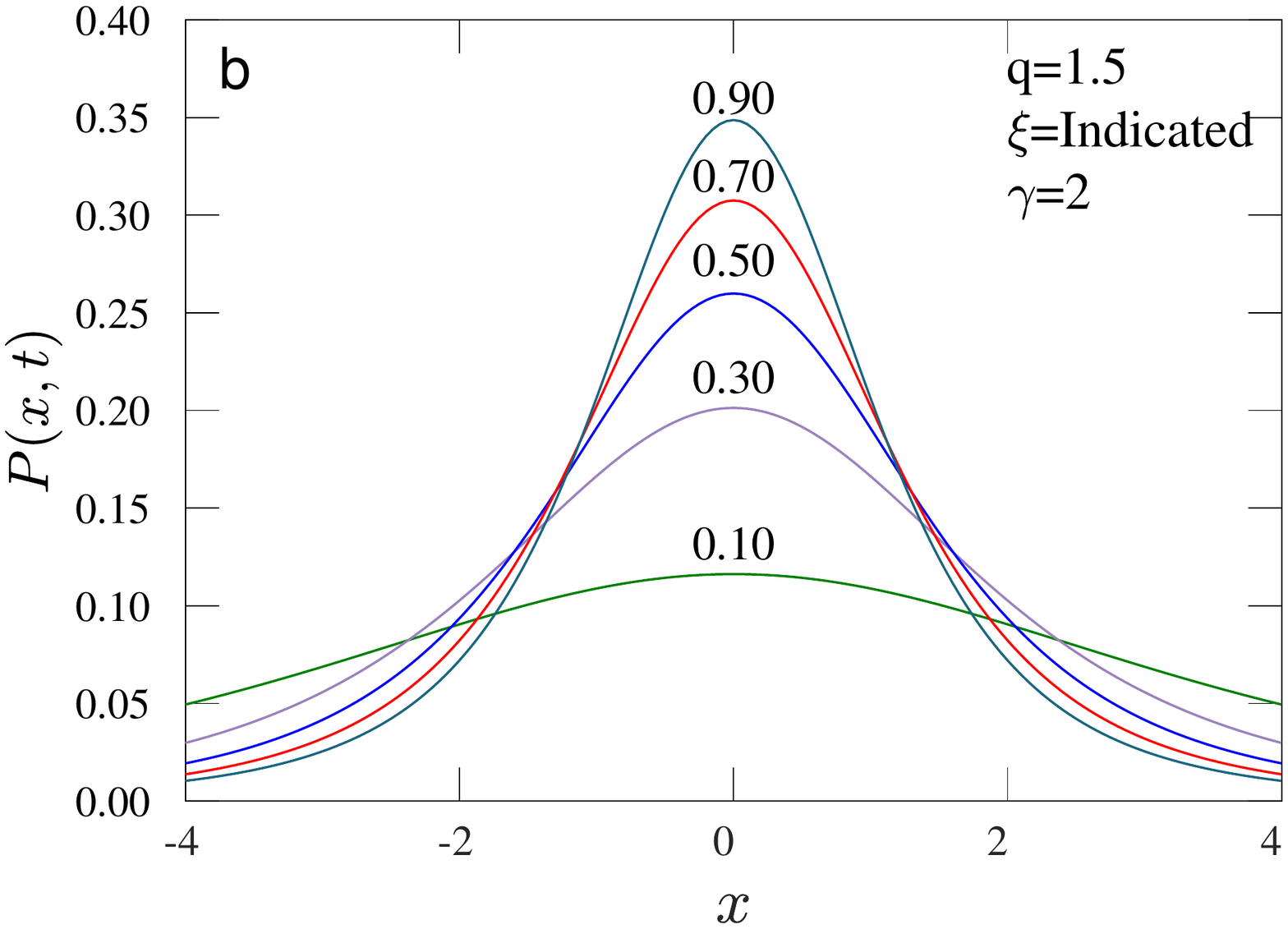}
\includegraphics[scale=0.40,trim=0cm 0cm 0cm 15cm]{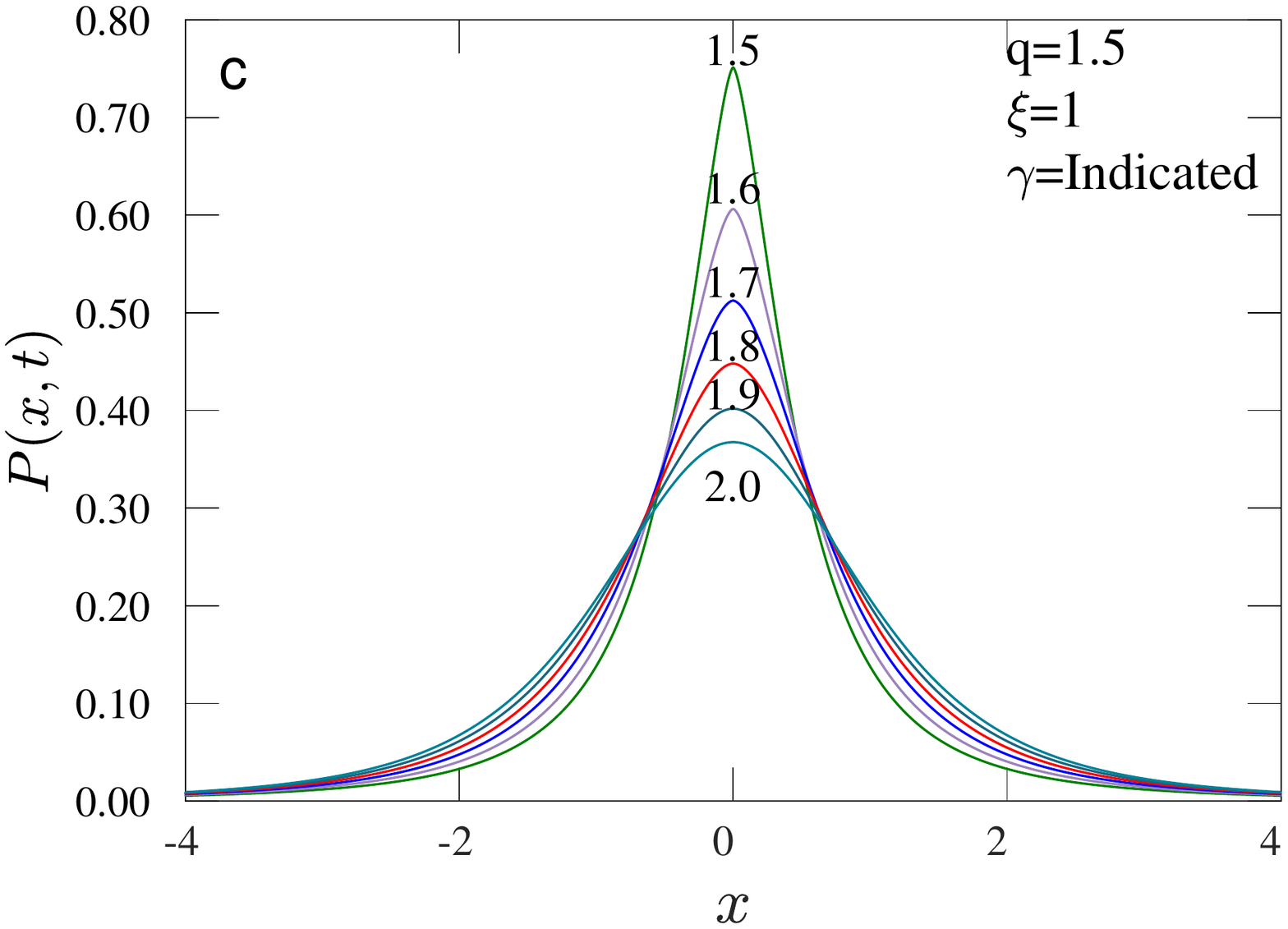}
\includegraphics[scale=0.40,trim=0cm 0cm 0cm 15cm]{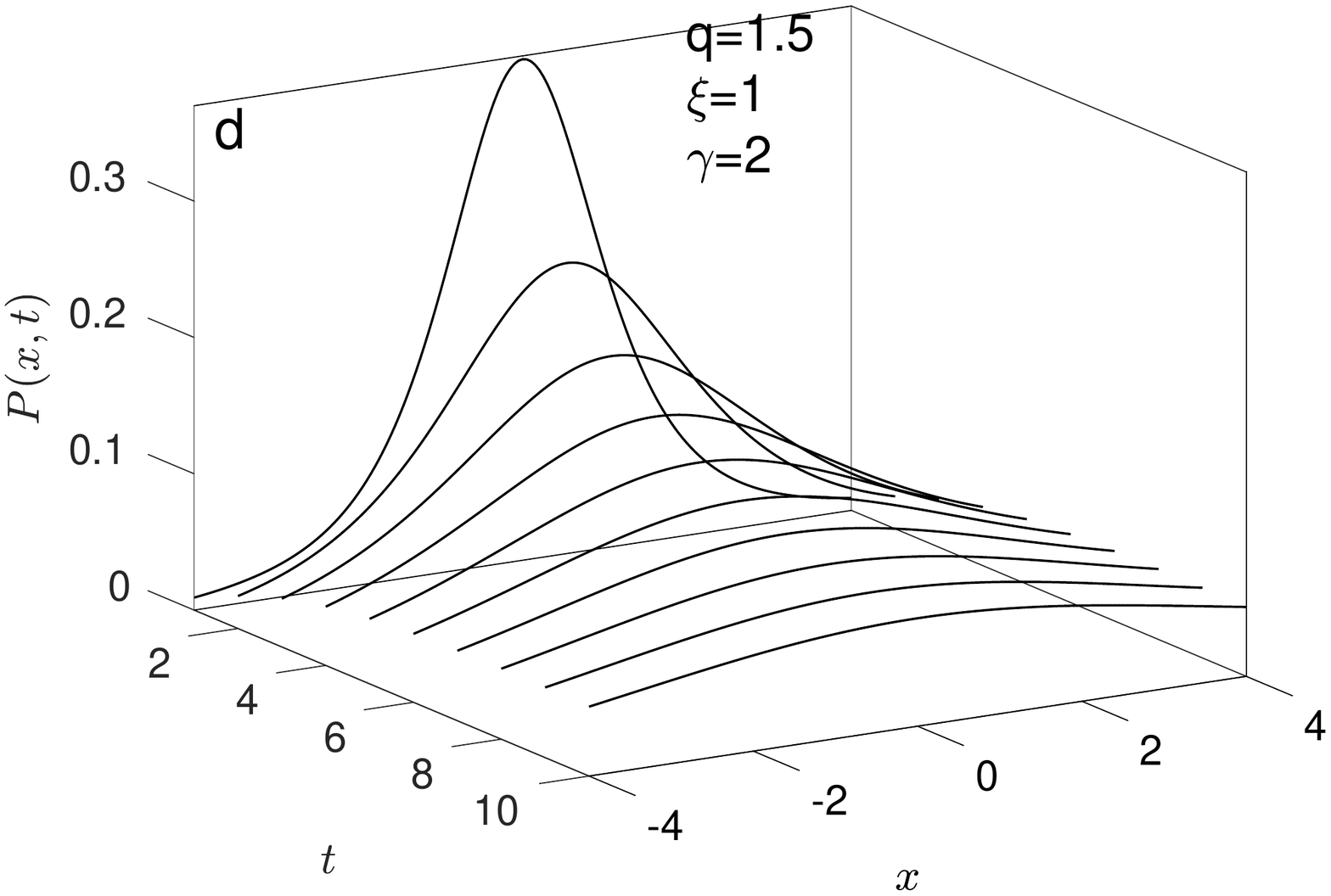}
\caption{(a) L$q$-Gaussians for different $q$ values as indicated in the plot for $ t=1, \xi=1, \gamma=2$ (b) L$q$-Gaussians for different $\xi$ values as indicated in the plot for $ t=1, q=1.5, \gamma=2$ (c) L$q$-Gaussians for different $\gamma$ values as indicated in the plot for $ t=1, q=1.5, \xi=1$ (d) Time evolution of the Green function of Eq.~\ref{eq:d_e_x}. Over time, the peak of curves decrease and the PDFs lost behavior of heavier tails, in contrast with (c), by increasing $\gamma,$ the peak of curves and behavior of heavier tails decrease.In (a), by increasing $q,$the peak of curves increase and the distribution becomes narrower (the tails become  heavier).  }
\label{fig:LqGaussian}
\end{figure*}
%%%%%%%%%%%%%%%%%%%%%%%%%%%%%%%%
 \subsection{A  local fractional nonlinear time-space diffusion equation}\label{SEC:localPME} 
In this section, we solved a time-space FPME (TS-FPME) with (LL) fractionalization. The Katugampola fractional definition was applied for  the time and space fractional operators. Such FPME can be written as: 
\begin{equation}
\begin{split}
&\dfrac{\partial^{\xi}}{\partial t^{\xi}}P(x,t)=D\dfrac{\partial^{\gamma}}{\partial x^{\gamma}}P^{\nu}(x,t),
\end{split}
\label{eq:d_e_x}
\end{equation}
where  $0<\xi \leq 1<\gamma \leq 2, $
$\vert\nu\vert<1$ are a set of three free parameters, and $D$ is the diffusion coefficient.  To solve Eq.~(\ref{eq:d_e_x}), we express the function $P(x,t)$ in its self-similar form: 
\begin{equation}
P(x,t)=\frac{1}{\phi(t)}F \left(\frac{x}{\phi(t)} \right),
\label{Eq:generalForm}
\end{equation}
where $\phi(t)$ is a function to be identified.
The Eq.~(\ref{Eq:generalForm}) is consistent with a  symmetric probability distribution.\par
By considering, $z=\frac{x}{\phi(t)}$, and  inserting Eq.~(\ref{Eq:generalForm}) into Eq.~(\ref{eq:d_e_x}), we have  the following two equations: 
\begin{equation*}
\begin{split}
&\partial^{\gamma}_{x}P^{\nu} = \frac{1}{\phi^{\nu+\gamma}}\frac{d^{\gamma}}{dz^{\gamma}}F^{\nu},\\
&\dfrac{\partial^{\xi}P}{\partial t^{\xi}}=\dfrac{-1}{\phi^{2}(t)}\dfrac{\partial^{\xi} \phi}{\partial t^{\xi}}\left[F+z\frac{d}{dz}F\right],
\end{split}
\end{equation*}
so that,
\begin{equation*}
\dfrac{-1}{\phi^{2}(t)}\dfrac{\partial^{\xi}\phi}{\partial t^{\xi}}\frac{d}{dz}[ zF]=\dfrac{D}{\phi^{\nu+\gamma}}\frac{d^{\gamma}}{dz^{\gamma}}F^{\nu}.
\end{equation*}
In the above equation, the properties of the Katugampola derivative were used  (see Appendix~\ref{SEC:Katugampola-derivative} and Table \ref{table:4}). Then, we arrange everything in such a way that all quantities in one side are only a function of $z$, and in the other side are a sole function of $t$. This procedure leads us to obtain the two following independent equations:
\begin{equation*}
\begin{split}
&\phi^{\nu+\gamma-2}\dfrac{\partial^{\xi}\phi}{\partial t^{\xi}}=\frac{\xi}{\nu+\gamma-1},\\
&-\dfrac{\xi}{\nu+\gamma-1}\frac{d}{dz}[ zF]=D\frac{d^{\gamma}}{dz^{\gamma}}F^{\nu}. 
\end{split}
\end{equation*}
The solution of the first equation is $ \phi\propto t^{\frac{\xi}{\nu+\gamma-1}}, $ and for the second one we have: 
\begin{equation*}
\frac{d}{dz}[ zF]=F+z\frac{d}{dz}F=k\frac{d^{\gamma}}{dz^{\gamma}}F^{\nu},
\end{equation*}
with $k=\frac{-D(\nu+\gamma-1)}{\xi}$, which can be rewritten as:
\begin{equation}
\frac{d}{dz}[ zF] =k\frac{d^{\gamma-1}}{dz^{\gamma-1}}\frac{d}{dz}[F^{\nu}].
\label{Eq:dF}
\end{equation}
By applying the  property: $\mathcal{D}^{\mu}(f)=\mathcal{D}^{\mu-1}\mathcal{D}^{1}(f)$ for $1<\mu\leqslant 2$, and taking local fractional integral with respect to $z$ in Eq.~(\ref{Eq:dF}), the following expression is obtained:
\begin{equation}
\int^{t}\frac{d}{dz}[ zF] \frac{dz}{z^{2-\gamma}}=k\int^{t}\frac{z^{2-\gamma}}{z^{2-\gamma}}\frac{d}{dz} \left[\frac{d}{dz} F^{\nu} \right]dz.
\label{Eq:integral}
\end{equation}
 In the right hand side, the integration by parts is used, $\frac{z^{\gamma-1}F}{\gamma-1}+c$, choosing $c=0$. By considering $ F=(c_{1}+c_{ 2}z^{\gamma})^{\frac{1}{\nu-1}}$ to obtain a special solution (where $c_1$ and $c_2$ are constants),  the following expression is obtained: 
\begin{equation*}
\frac{d }{dz }[F^{\nu}]=\frac{\gamma \nu c_{ 2}}{\nu-1}z^{\gamma-1}F.
\end{equation*}
After incorporate the previous expression into the Eq.~(\ref{Eq:integral}), we obtain:
\begin{equation*}
c_{ 2}=\dfrac{-(\nu-1)\xi}{D\gamma (\gamma-1)\nu(\nu+\gamma-1) }.
\end{equation*}
Therefore, the general solution is:
\begin{equation}
P(x,t)\propto\frac{1}{t^{\frac{\xi}{\nu+\gamma-1}}} \left(c_{ 1}+\dfrac{(\nu-1)\xi}{D\gamma(\gamma-1) \nu (\nu+\gamma-1)}\frac{x^{\gamma}}{t^{\frac{\gamma\xi}{\nu+\gamma-1}}} \right)^{\frac{1}{\nu-1}},
\label{Eq:P_x_t_l}
\end{equation}
 where $c_1$ is removed after apply the normalization condition in Eq.~(\ref{Eq:P_x_t_l}). Then, by defining  $\nu=2-q$, considering $ \alpha=\frac{1-q+\gamma}{\xi}$, and
\begin{equation}
\begin{split}
\eta^{L}_{q}(\xi,\gamma,q,D)&=\dfrac{\xi}{D\gamma (\gamma-1)\nu(\nu+\gamma-1)}\\
&=\dfrac{\xi}{D\gamma(\gamma-1)(2-q)(1-q+\gamma)},
\end{split}
\end{equation}
the following equation is reached:
\begin{equation}
P(x,t)=\frac{A^{L}_{q}}{t^{\frac{1}{\alpha}}}[1+(1-q)\eta^{L}_{q}(\beta,\gamma,q,D)\frac{x^{\gamma}}{t^{\frac{\gamma}{\alpha}}}]^{\frac{1}{1-q}},
\label{eq:p_x}
\end{equation}
 where $A^{L}_{q}$ is a normalization factor. In most of  physical systems cases, the $P(x,t)$ is symmetric with respect to $x$. This point leads us to make $x\rightarrow |x|$ (i.e. its absolute value), or we can consider some values of $\gamma$ that satisfy this property.
 \\
 Then, the normalization factor was identified as follows:
\begin{equation}
A^{ L}_{q}= \frac{1}{2}\left( \eta(\xi,\gamma,q,D)\right)^{\frac{1}{\gamma}} \dfrac{\Gamma(\frac{ 1}{q-1})}{\Gamma(\frac{1}{q-1}-\frac{1}{\gamma})\Gamma(1+\frac{1}{\gamma})}, 
\label{Eq:prefactorLocal}
\end{equation}
where
\begin{equation*}
\eta(\xi,\gamma,q,D)=\dfrac{ \eta^{ L}_{q}(\beta,\gamma,q,D)}{(\nu-1)^{-1}}.
\end{equation*}
For $ \gamma=2$, these parameters become $\alpha=\frac{3-q}{\xi}$, $A^{ L}_{q}=\sqrt{\pi^{-1}\eta(\xi,q,D)} \dfrac{\Gamma(\frac{ 1}{q-1})}{\Gamma(\frac { 3-q} {2(q-1)})}$, and $\eta^{L}_{q}(\xi ,q,D)= \dfrac{\xi}{2D (2-q)(3-q)}$. \\

 We call  Eq.(\ref{eq:p_x}) the local $q$-Gaussian (L$q$-Gaussian) distribution, which is the Green function of Eq.(\ref{eq:d_e_x}) obtained from a TS-FPME  with the Katugampola fractional derivative (local fractional definition). The L$q$-Gaussian has been defined as, $g_{q}^{\gamma}(x)$, Equation N.$6$ in Table \ref{table:1}. \\

In Figure ~\ref{fig:LqGaussian}\textit{,} we show the L$q$-Gaussians for different $q$ values as indicated in the plot for  $ t=1, \xi=1, \gamma=2$. In subfigure ~\ref{fig:LqGaussian}a,  by increasing $q,$ the peak of curves increase and the distribution becomes narrower  (the tails become heavier).  In the case of subfigure~\ref{fig:LqGaussian}b something similar occurs, where we denote the PDFs of L$q$-Gaussian for different $\xi$ values as indicated in the plot for $ t=1, q=1.5$ and $\gamma=2$. The reverse occurs for   Figure~\ref{fig:LqGaussian}c, by increasing  $\gamma$ (considering $ t=1, q=1.5, \xi=1$), the peak of the   PDFs  decrease. Also,   the time evolution of the Green function of Eq.~\ref{eq:d_e_x}  is shown in subfigure~\ref{fig:LqGaussian}d.

%(KARINA, L-Q GAUSSIAN ARE THE GENERALIZED GAUSSIAN YOU SHOW IN THE TABLE. PLEASE CLARIFY THAT)

%%%%%%%%%%%%%%%%%%%%%%%%%%
\subsection{A connection between the $(q,\alpha)$-stable distributions and L$q$-Gaussians}
In the recent subsection, we obtained   L$q$-Gaussian distributions by solving the TS-FPME. In fact, these L$q$-Gaussians are a  generalized  $q$-Gaussians by considering  $|x|^{\gamma/2},~q>1$, i.e., a $q-$exponential in the variable $|x|^{\gamma}.$   From the definition of the $q$-exponential, it follows that  $f \sim C_{f}|x|^{- \gamma /( q-1 )},\,\,C_{f}>0, $ as $|x|\rightarrow\infty.$ Analogously, for any  $q$-Gaussian, $g \sim C_{g}|x|^{- 2 /( q-1 )},\,\,C_{g}>0, $ as $|x|\rightarrow\infty$.  By comparing the order of the power law of the asymptotes,   we verify that for a fixed  $1<\gamma<2$, and for
any $1<q<2$ there exists a  proportionality  from L$q$-Gaussians to $q$-Gaussian. For further details, see ~\cite{Umarov}. 

Let us denote the class of random variables with $(q,\gamma)$-stable distributions by $\mathcal{L}_{q}[\gamma].$ A random variable $X\in\mathcal{L}_{q}[\gamma]$ has a symmetric density $f(x)$  with asymptotes  $f \sim C|x|^{-(1+\gamma)/(1+\gamma(q-1))},\,|x|\rightarrow\infty,$ where $1\leq q<2,\,1<\gamma<2,$ and $C$ is a positive constant.
On the other hand, any L$q$-Gaussian behaves asymptotically  when  $C_{1}/|x|^{\gamma/q-1}$. Especially any L$q_{\gamma}$-Gaussian behaves asymptotically  when  $C_{2}/|x|^{\gamma/(q_{\gamma}-1)}.$ Hence, we obtain the following   relationship :
\begin{equation}
\dfrac{1+\gamma}{1+\gamma(q-1)}=\dfrac{\gamma}{q_{\gamma}-1}.
\end{equation}
Solving this equation with respect to $q_{\gamma},$ we have
\begin{equation}\label{stable_Gaussian}
q_{\gamma} =\dfrac{\gamma Q_{\gamma}+1 }{ \gamma+1},\,\,\,Q_{\gamma}=2+\gamma(q-1).
\end{equation}
  Three parameters were linked:  $\gamma,$ the parameter of the $\gamma$-stable Levy distributions, $q,$ the parameters of correlations, and $q_{\gamma},$ the parameters of attractors in terms of L$q_{\gamma}$-Gaussians.  Then under Eq.~(\ref{stable_Gaussian}) the density of $X\in\mathcal{L}_{q}[\gamma]$ is asymptotically equivalent to L$q_{\gamma}$-Gaussian.
\\
The  L$q$-Gaussians have an  interesting property.  Its successive  derivatives, and integrations with respect to $|x|^{\gamma }$ correspond to $q_{\gamma,n}-$exponentials in the same variable $|x|^{\gamma}$, 
where $q_{\gamma,n}=\frac{\gamma q-n(q-1)}{\gamma-n(q-1)} $~\cite{Umarov}.\\
 By considering  $\mathcal{G}_{q_{\gamma,n}}[\gamma]$ as a set of functions $\lbrace be_{q_{\gamma,n}}^{-\beta|\xi|^{\gamma}},b>0, \beta>0\rbrace,$ and $\mathcal{F}_{q_{\gamma,n}}$ be the $ {q_{\gamma,n}}-$Fourier transform,   the following expression is obtained: 
\begin{equation*}
\mathcal{F}_{q_{\gamma,n}}:\mathcal{G}_{q_{\gamma,n}}[\gamma]\rightarrow\mathcal{G}_{q_{\gamma,n+1}}[\gamma],\,\,\,-\infty<n\leq [\gamma/(q-1)].
\end{equation*}
This is similar to the $q-$exponential with the variable $|\xi|^{\gamma} $, i.e. $e_{q}^{-\beta|\xi|^{\gamma}},\,\beta>0,$  which is the $q-$Fourier transform of $(q,\gamma)-$stable distributions~\cite{Umarov}.

%%%%%%%%%%%%%%%%%%%%%%%%%%
\subsection{The local fractional nonlinear time-space diffusion equation with the drift}
The drift  is often an inevitable part of stochastic systems , that should be analyzed in details for every case study to control its effects. Although it is  suggested to define the equations for the general drift term. For the case where it depends only on time (as the case for many physical systems of interest), the situation becomes easier. In this case, the governing equation is:
\begin{equation}\label{general eq}
\begin{split}
\dfrac{\partial^{\xi}}{\partial t^{\xi}}&P(x,t)=-a(t)\dfrac{\partial P}{\partial x}+ D\dfrac{\partial^{\gamma} P^{\nu}}{\partial x^{\gamma}},\\
& 0<\xi \leq 1< \gamma\leq 2,\ \  \vert\nu\vert < 1.
\end{split}
 \,\,\,\,\,  
\end{equation}
By change of variable $\tau=  t ^{\xi}$ and the definition of the Katugampola derivative, we have:
\begin{equation*}
\partial_{\tau}P =- a^{\prime}  (\tau)\partial_{x}P +D\partial_{x}^{\gamma}P^{\nu},
\end{equation*}
where $ a^{\prime}  (\tau)=\frac{1}{\xi} a(t(\tau))$. By using the change of variable $(s,y)\equiv (\tau,x-x_{0}-f(\tau))$, where $f(\tau)=\int_{0}^{\tau} a^{\prime}  (  \tau^{\prime} )d \tau^{\prime} $, and using the fact that $\frac{\partial y }{\partial \tau }=-a^{\prime}(\tau)$ and $\partial_{\tau}+a^{\prime}  (\tau)\partial_{x}=\partial_{s}$, one finds that  the governing equation  $P(y,\tau)$ is:
\begin{equation*}
\partial_{\tau}P(y,\tau)=D\partial_{y}^{\gamma}P^{\nu}(y,\tau),
\end{equation*}
for which the solution is $(x_{0}\equiv 0),$ 
\begin{align*} 
P&(y,\tau)=\frac{A}{\tau^{\frac{1}{1+\gamma-q}}}\\
& \times(c_{1}-\dfrac{1-q}{2D\gamma(\gamma-1)(2-q)(1+\gamma-q)}\frac{y^{\gamma}}{\tau^{\frac{\gamma}{1+\gamma-q}}})^{\frac{1}{1-q}}.
\end{align*}
 Let us equate  the $P(y,\tau)$: 
\begin{equation*}
P(x,t)=\frac{\partial {y}}{\partial {x}}P(y,\tau( t ) ).
\end{equation*}
 Then, we obtain that:
\begin{align}\nonumber
P&(x,t)=\frac{A}{t^{\frac{\xi}{1+\gamma-q}}}\\
& \times \left( c_{1}-\dfrac{1-q}{2D\gamma(\gamma-1)(2-q)(1+\gamma-q)}\frac{(x-f^{\prime}(t))^{\gamma}}{t^{\frac{\gamma\xi}{1+\gamma-q}}} \right)^{\frac{1}{1-q}},
\label{Eq:LPME_drift}
\end{align}
where $f^{\prime}(t)=f(\tau( t )),\,\,A$  is a normalization factor, and $c_{1}$ is a constant. The Eq.~(\ref{Eq:LPME_drift}) is a L$q$-Gaussian solution with a drift.
%%%%%%%%%%%%%%%%%%%%%%%%%%%%%%%%%%%%%

\section{PME with Local and Non-local Fractional Operators}~\label{SEC:L-N}
To be self-contained, we consider the case where one fractional derivative is local, and the other is non-local. Therefore,  in this section  we solve  a TS-FPME   with (LN)   fractionalization ,  
\begin{equation}
\begin{split}
&\dfrac{\partial^{\xi}}{\partial t^{\xi}}P(x,t)=D\dfrac{\partial^{\gamma}}{\partial x^{\gamma}}P^{\nu}(x,t),\\
& 0<\xi \leq 1,\,\,0<\gamma<\frac{1}{2},\,\, \nu>-1,
\end{split}
\label{ufo10}
\end{equation}
where $\dfrac{\partial^{\xi}}{\partial t^{\xi}}$ and $\dfrac{\partial^{\gamma}}{\partial x^{\gamma}}$ denote  the Katugampola and Riemann-Liouville fractional derivatives, respectively (see Table \ref{table:3}).
We consider the decomposition  of  (Eq.~\ref{Eq:generalForm}). Then, by using the property of $\frac{d^{\gamma}}{dx^{\gamma}}F(ax)=a^{\gamma}\frac{d^{\gamma}}{dz^{\gamma}}F(z)$, and some properties of Katugampola derivative we obtain:
\begin{equation*}
\partial^{\gamma}_{x}P^{\nu} = \frac{1}{\phi^{\nu+\gamma}}\frac{d^{\gamma}}{dz^{\gamma}}F^{\nu},
\end{equation*}
\begin{equation*}
\dfrac{\partial^{\xi}P}{\partial t^{\xi}}=\dfrac{-1}{\phi^{2}(t)}\dfrac{\partial^{\xi}\phi}{\partial t^{\xi}}\left[F+z\frac{d}{dz}F\right],
\end{equation*}
so that,
\begin{equation*}
\dfrac{-1}{\phi^{2}(t)}\dfrac{\partial^{\beta}\phi}{\partial t^{\xi}}\frac{d}{dz}[ zF]=\dfrac{D}{\phi^{\nu+\gamma}}\frac{d^{\gamma}}{dz^{\gamma}}F^{\nu}.
\end{equation*}
   To continue,  similar strategies  applied in Sec.~\ref{SEC:localPME} were used. We transform the previous equation  into two independent equations:
\begin{equation*}
\phi^{\nu+\gamma-2}\dfrac{\partial^{\xi}\phi}{\partial t^{\xi}}=\frac{\xi}{\nu+\gamma-1},
\end{equation*}

\begin{figure*}[!htbp]
\centering 
\includegraphics[scale=0.40,trim=0cm 0cm 0cm 15cm]{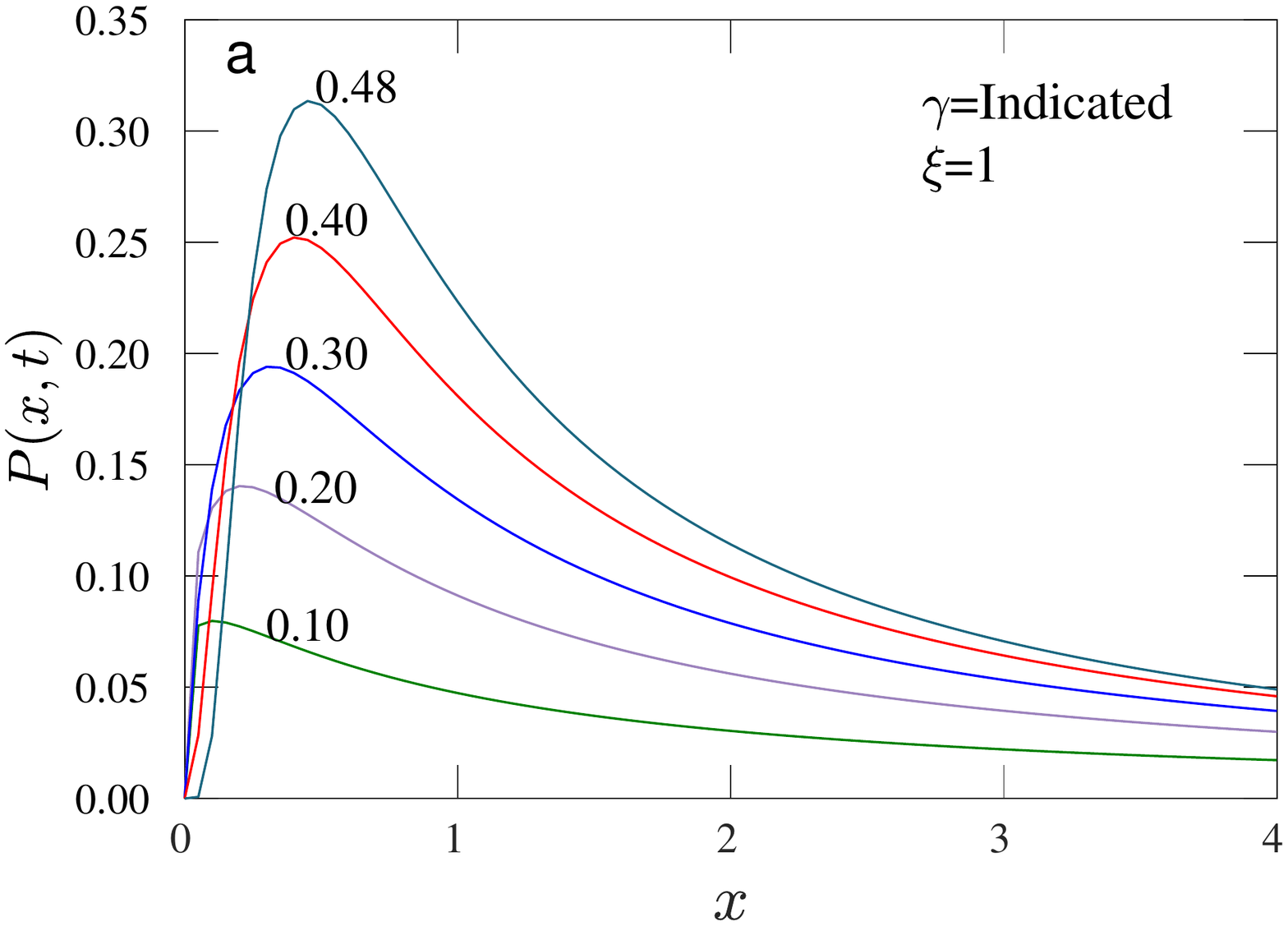}
\includegraphics[scale=0.40,trim=0cm 0cm 0cm 15cm]{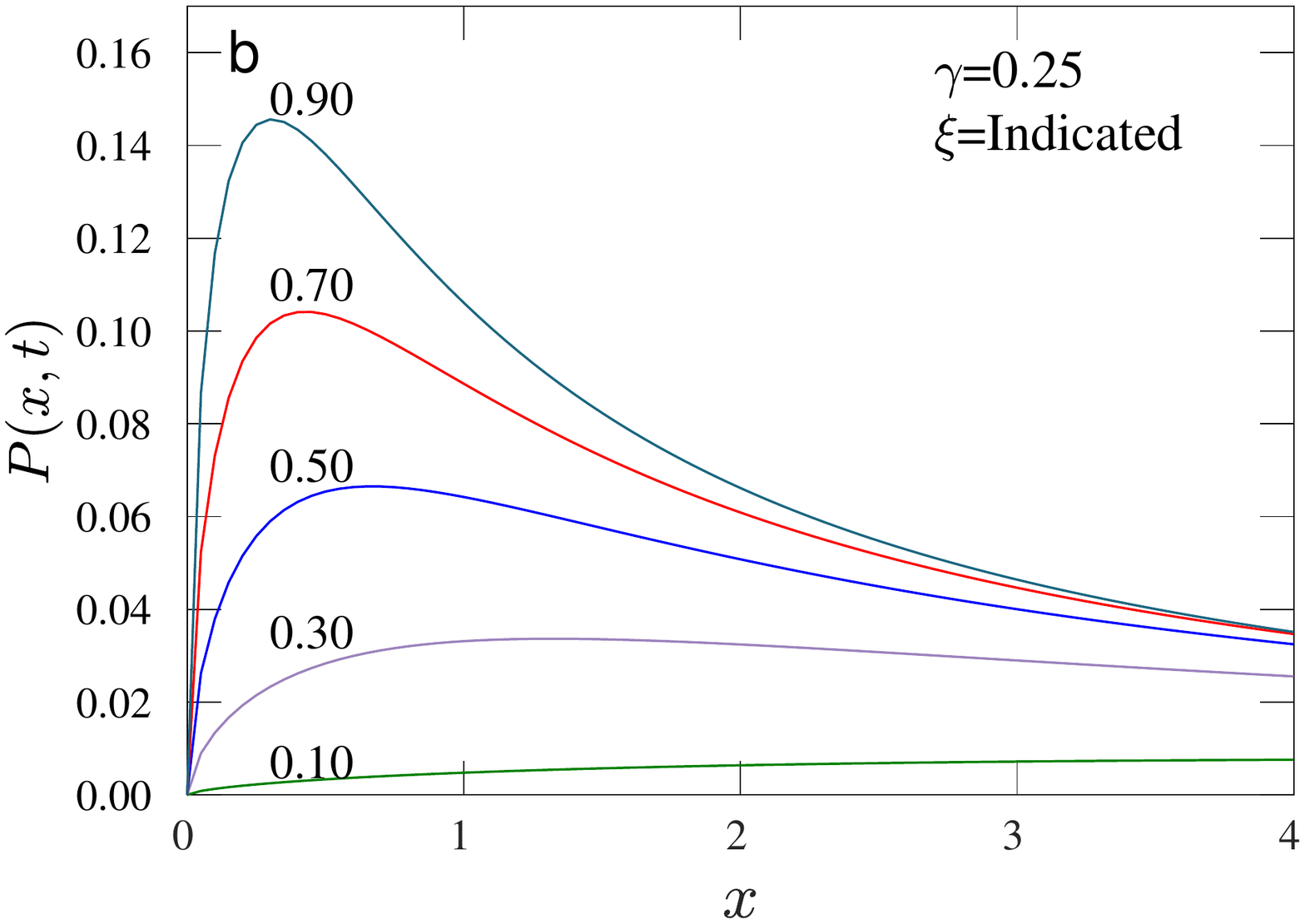}
\includegraphics[scale=0.40,trim=0cm 0cm 0cm 15cm]{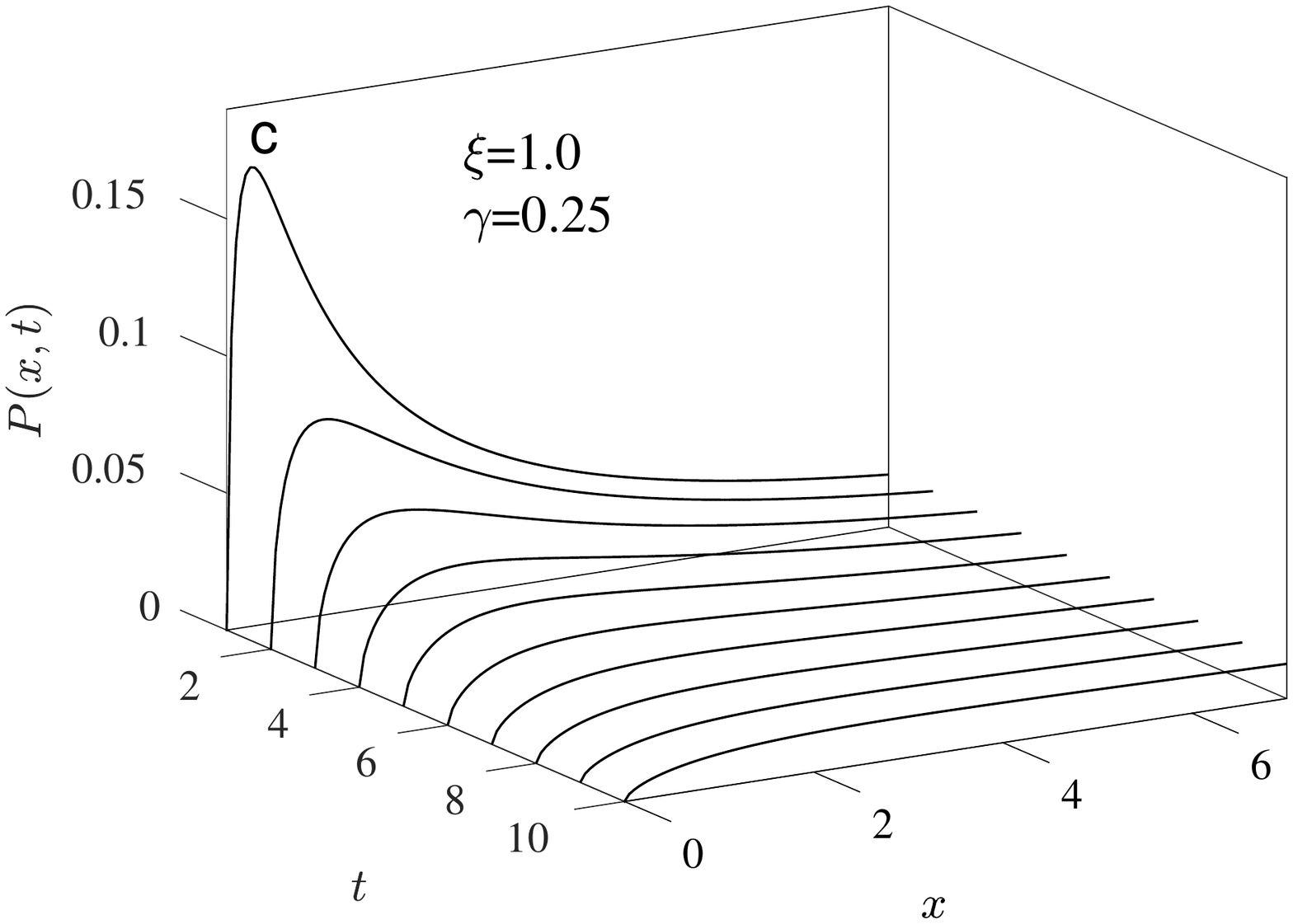}
\caption{(a)The PDFs of the LNL$q$-Gaussian distributions for different values of $ \gamma $ and $ \xi=1$. (b)The PDFs of the LNL$q$-Gaussian distributions for different values of $ \xi $ and $ \gamma=0.25$. (c) Time evolution of the Green function of Eq.~\ref{ufo10} . }
\label{fig:NLLqGaussian}
\end{figure*}

\begin{equation*}
-\dfrac{\xi}{\nu+\gamma-1}\frac{d}{dz}[ zF]=D\frac{d^{\gamma}}{dz^{\gamma}}F^{\nu}, 
\end{equation*}
where the solution of the first equation is $ \phi=t^{\frac{\xi}{\nu+\gamma-1}}$.
For the second,  the solution is : 
\begin{equation*}
\frac{d}{dz}[ zF]=F+z\frac{d}{dz}F=-D\left[\frac{\nu+\gamma-1}{\xi}\right]\frac{d^{\gamma}}{dz^{\gamma}}F^{\nu},
\end{equation*}
or $\frac{d}{dz}[ zF]=k\frac{d^{\gamma}}{dz^{\gamma}}F^{\nu} $  with $k=\frac{-D(\nu+\gamma-1)}{\xi}.$ 
By  integrating  with respect to $z,$ we  obtain :
\begin{equation}\label{ufo9}
zF=k\frac{d^{\gamma-1}}{dz^{\gamma-1}}[F^{\nu}]+c,
\end{equation}
 where $c$ is a constant,we have set $c=0$.  By considering  $F(z)=z^{\mu}(c_{1}+c_{2}z)^{\lambda}$,   and using   
the following property for the RL operators,
\begin{equation}\label{Eq:RLderivative}
\mathcal{D}_{x}^{\delta}[x^{\alpha}(a+bx)^{\beta}]=a^{\delta}\dfrac{\Gamma(\alpha+1)}{\Gamma(\alpha+1-\delta)}x^{\alpha-\delta}(a+bx)^{\beta-\delta},
\end{equation}
we find that 
\begin{align*}
\frac{d^{\gamma-1}}{dz^{\gamma-1}}[F^{\nu}]=c_{1}^{\gamma-1}\frac{\Gamma(\nu\mu+1)}{\nu\mu+2-\gamma}z^{\nu\mu-\gamma+1}(c_{1}+c_{2}z)^{\nu\lambda-\gamma+1}.
\end{align*}
We put this result into Eq.(\ref{ufo9}), and  the following expressions are obtained: 
\begin{equation*}
\nu=\dfrac{2-\gamma}{1+\gamma},\,\,\mu=\dfrac{\gamma(1+\gamma)}{1-2\gamma},\,\,\lambda=\dfrac{ 1-\gamma^{2}}{ 2\gamma-1},
\end{equation*}
\begin{equation*}
c_{1}^{\gamma-1}=\dfrac{ -\beta(1+\gamma) \Gamma((2+\gamma(\gamma-3))(1-2\gamma)^{-1})}{D(1+\gamma(\gamma-1))\Gamma((1-\gamma^{2})(1-2\gamma)^{-1})  }.
\end{equation*}
 The recent expressions  reveal  that the master equation admits the following solution:
\begin{equation}\label{ufo11}
P(x,t)=\frac{A}{t^{\frac{1}{\sigma}}}\left(\frac{x}{t^{\frac{1}{\sigma}}}\right)^{\alpha\gamma}\left(c_{1}+c_{2}\frac{x}{t^{\frac{1}{\sigma}}}\right)^{-\alpha(1-\gamma)},
\end{equation}
where $\alpha=\frac{1+\gamma}{1-2\gamma},\,\sigma=\frac{1-\gamma(1-\gamma)}{\beta(1+\gamma)},\,c_2=1 $ and
\begin{equation*}
A= \dfrac{\Gamma(\alpha(1-\gamma))}{c_{1}^{\gamma }\Gamma(1+\alpha\gamma)\Gamma(\gamma)}.
\end{equation*}
 This solution is named local-non-local (LNL) $q$-Gaussian distribution.  The solution is not a generalized $q$-Gaussian distribution. This occurs due to the factor, $x^{\alpha\gamma}$, which is located in front of the solution assuring that this class of PDFs tend to zero when $x\rightarrow 0$. This makes a great difference with the previous function, the L$q$-Gaussian Eq.~(\ref{eq:p_x}). \\

Figure~\ref{fig:NLLqGaussian} shows the LNL$q$-Gaussian distribution functions. All the graphs tend to zero as $x\rightarrow 0$ for all times, as stated above, and don't have symmetric form around their peaks in terms of $x$. They  manifested  an abrupt increase before the peak, and long-range tails beyond it, with peaks depending on $q$ and $\xi$. The peak values increase by increasing both $\gamma$ and $\xi$.

%%%%%%%%%%%%%%%%%%%%%%%%%%%%%%%%
\begin{figure*}[!htbp]
\centering  
\includegraphics[scale=0.40,trim=0cm 0cm 0cm 15cm]{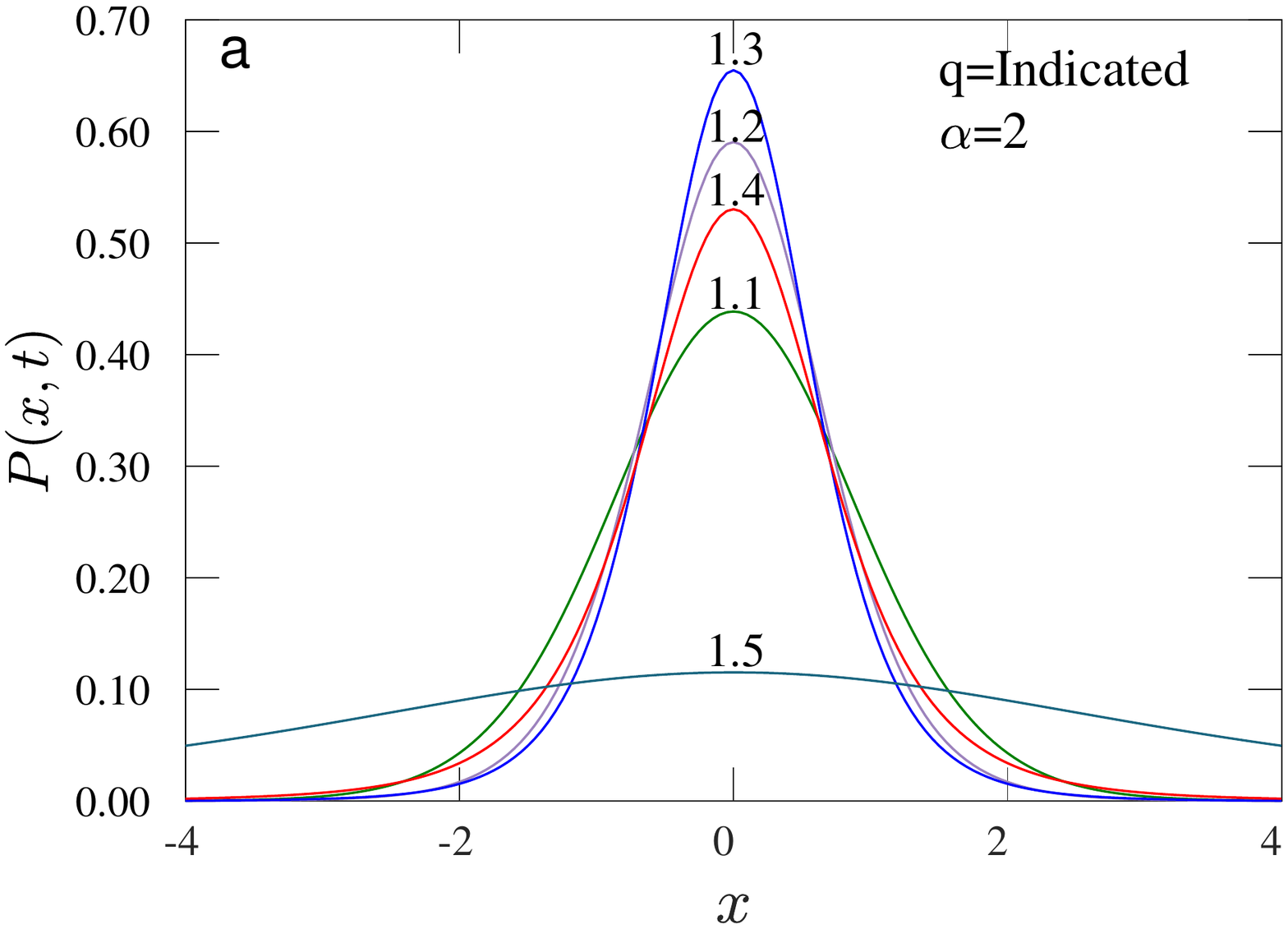}
\includegraphics[scale=0.40,trim=0cm 0cm 0cm 15cm]{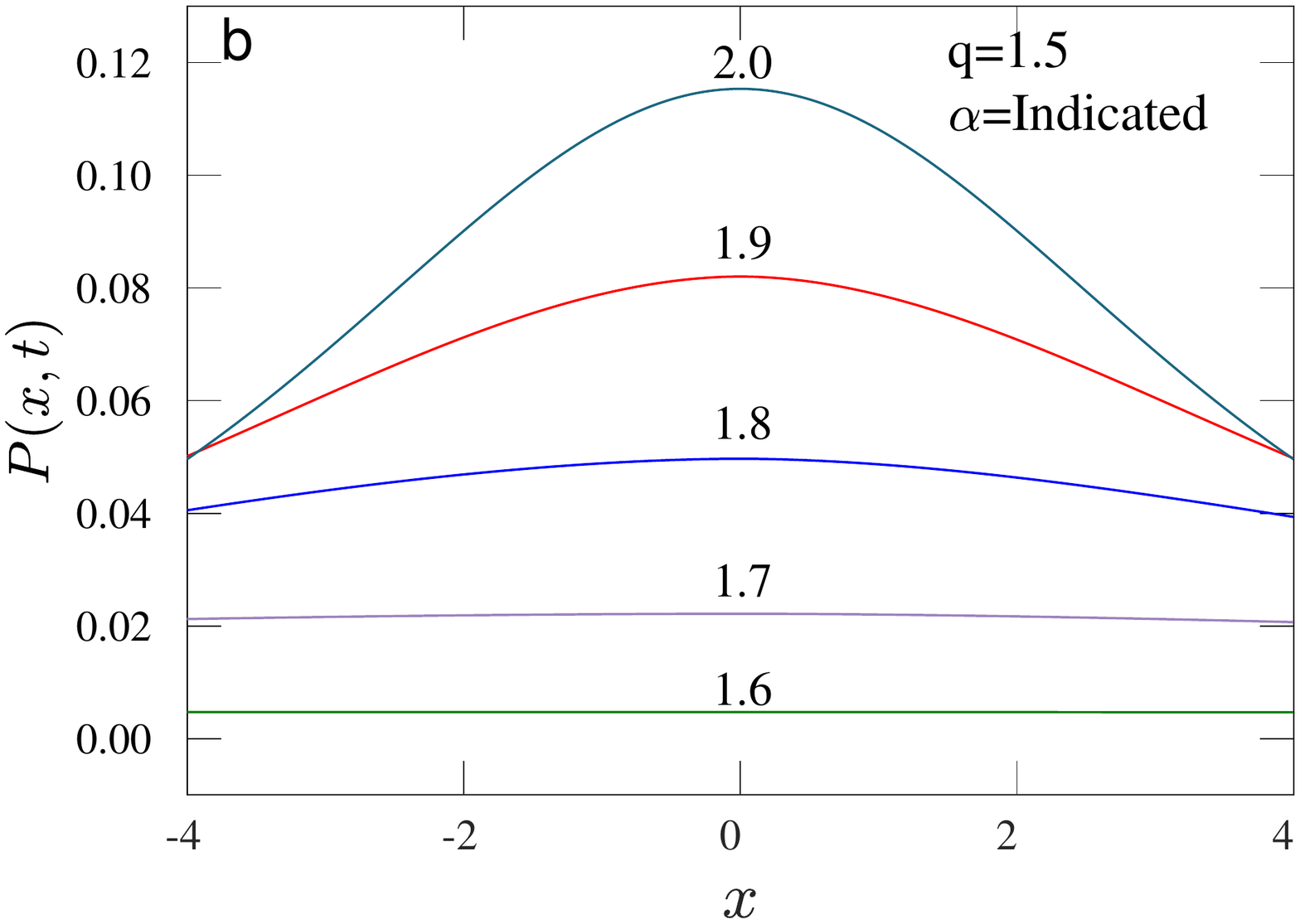}
\includegraphics[scale=0.40,trim=0cm 0cm 0cm 15cm]{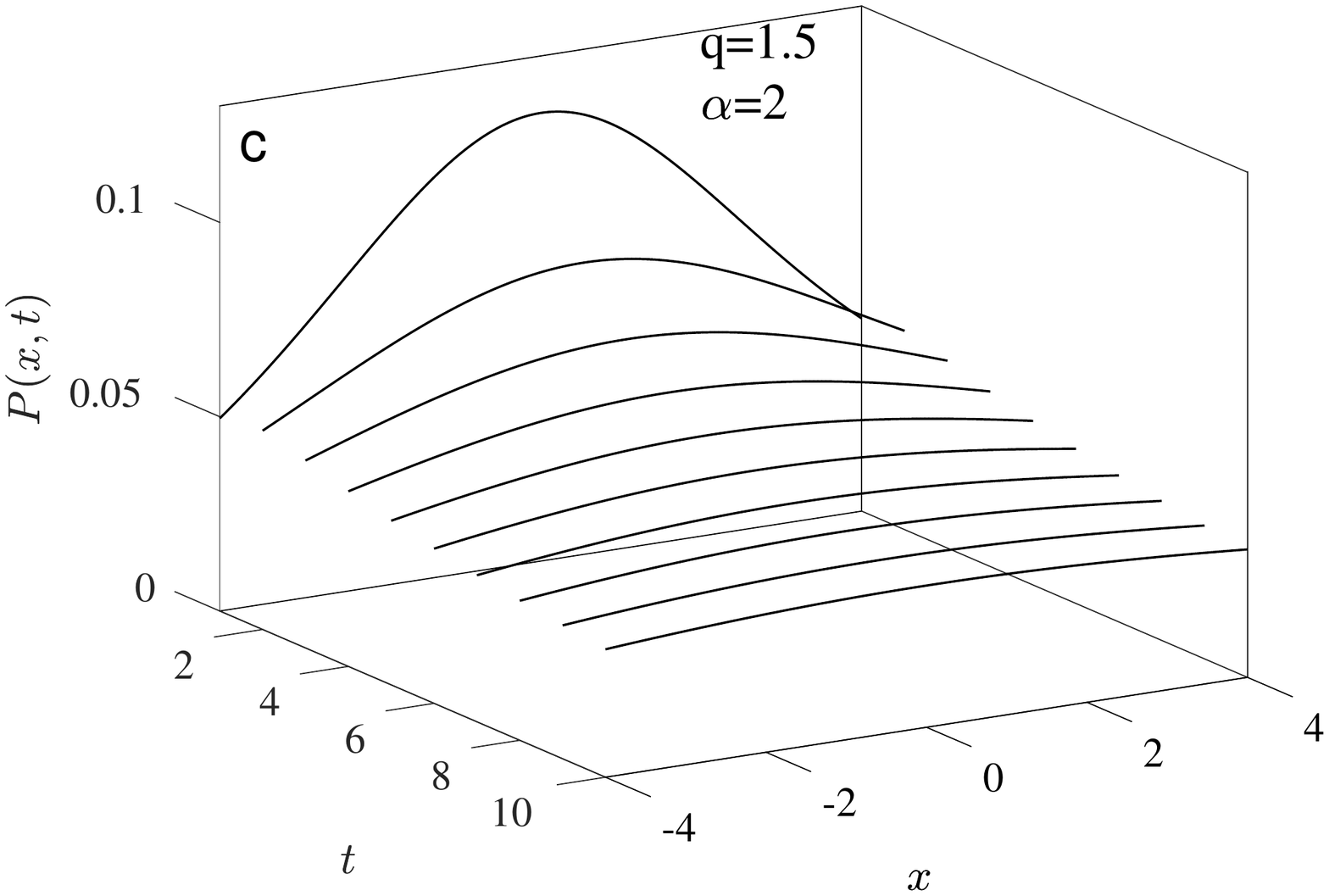}
\caption{(a)The PDFs of the  NL$q$-Gaussian distributions for  $ \alpha  =2$ and indicated values of $q$  (b)The PDFs of the  NL$q$-Gaussian distributions for $q=1.5$ and indicated value of $\alpha$. (c) The time evolution of PDFs of the NL$q$-Gaussian distributions for $q=1.5$ and $\alpha=2 $.  These are the Green functions of Eq.~(\ref{ufo1}). We can see, the NL$q$-Gaussian for a constant value $\alpha $ and different $q$ values shows a multiple behavior; for $q=1.1, 1.2, 1.3 $ the peak of curves and behavior of heavier tails increase, in contrast with $q=1.4, 1.5. $ Also, for a constant value $q $ and different $\alpha$ values, by increasing $\alpha,$ the peak of curves and behavior of heavier tails increase. In time evolution of L$q$Gaussian, over time, the peak of curves decreases and the PDFs lost behavior of heavier tails.}
\label{fig:NLqGaussian}
\end{figure*}
\section{PME with Non-local Fractional Operators}\label{SEC:N-N}
In this section we  solve one particular case of the G-PME displayed in Eq.(\ref{Eq:Matrix}), when $n=-1$, $a=0$, and $b=1$. The solution is obtained by considering a hybrid case, where the time derivative is RL operator, and the space derivative is Caputo operator (Appendix~\ref{SEC:Caputo-derivative}).   The other cases (RL-RL, Caputo-Caputo, and Caputo-RL derivatives) are straightforward to be processed following the same lines as this study. The resulting  FPME equations is
\begin{equation}\label{ufo1}
_{0}\mathcal{D} ^{\xi,\frac{1}{t}}P(x,t)= D \left( ^{C}\mathcal{D}_{1}^{\gamma, x^{\alpha}}P^{\nu}(x,t) \right),\,\,\,\,\,\,\,\xi,\gamma>0. 
\end{equation}
where $ \mathcal{D} $ is the RL operator for time derivative and $ ^{C}\mathcal{D} $ is the Caputo operator acting on ``space'' coordinate $x$. To construct our solution, we need to restrict ourselves to the case $ \nu=\frac{1+\xi}{1-\xi} $ leaving two parameters free for fitting, $\gamma$ and $\alpha$.\\

We again search for the solutions of  the form of  Eq.~(\ref{Eq:generalForm}), where the parameters were defined in the Section \ref{SEC:local-local} . In the following we show that the above equation admits the solution of the form $F(z)=(1+bz^{\alpha})^{\lambda}$, where $ b\neq0$, and $\alpha\neq 1$. When stablished, this solution serves as another variant of the generalized $q$-Gaussian solution.  By inserting  this form in the right side of Eq.~(\ref{ufo1}) we obtain:
\begin{eqnarray}\label{ufo3} \nonumber
^{C} \mathcal{D}_{1}^{\gamma, x^{\alpha}}P^{\nu}(x,t)&=&\left(\frac{1}{t^{\frac{1}{\alpha}}}\right)^{\nu}~~^{C}  \mathcal{D}_{1}^{\gamma, x^{\alpha}}(1+bz^{\alpha})^{\lambda\nu}
\\ 
=b^{\gamma}\left(\frac{1}{t}\right)&^{\frac{ \nu}{\alpha}+\gamma}&\dfrac{\Gamma(\lambda\nu+1)}{\Gamma(\lambda\nu+1-\gamma)}(1+bz^{\alpha})^{\lambda\nu-\gamma}.
\end{eqnarray}
 To obtain  the above equation, we have used the following property of the fractional derivatives, that is valid for all the fractional differential operators considered here,
\begin{equation*}
\mathcal{D}^{\mu}_{t}[f(at)]=a^{\mu}\mathcal{D}^{\mu}_{x}[f(x)]\mid_{x=at}.
\end{equation*}
 Additionally,  the following property of the fractional derivative of a function \textit{with respect to
another function}~\cite{ref19} was applied :
\begin{equation}
\begin{split}
 \mathcal{D}_{b}^{\alpha, \psi}(\psi(b)-\psi(x))^{\beta-1}&=\dfrac{\Gamma(\beta)}{\Gamma(\beta-\alpha)}(\psi(b)-\psi(x))^{\beta-\alpha-1},\\
&\alpha>0,\, n<\beta\in\mathbb{R}. 
\end{split}
\end{equation}
where $ n=\alpha, $ if $ \alpha\in \mathbb{N}$ and $ n=[\alpha]+1, $ if $ \alpha\notin \mathbb{N}.$\\
 By inserting  Eq.~(\ref{Eq:generalForm}) in the left side of Eq.~(\ref{ufo1}) and then 
  applying  Eq.~(\ref{Eq:RLderivative})  for the RL operators 
with $ \delta=\alpha+\beta+1,$ we get  
\begin{eqnarray}\label{ufo4}\nonumber
_{0}\mathcal{D} ^{\xi,\frac{1}{t}}P(x,t)&=&_{0}\mathcal{D} ^{\xi,\frac{1}{t}}\lbrace \frac{1}{t^{\frac{1}{\alpha}}}(1+bz^{\alpha})^{\lambda}\rbrace\\ \nonumber
&=&_{0}\mathcal{D} ^{\xi,\frac{1}{t}}\lbrace \frac{1}{t^{\frac{1}{\alpha}}}(1+b'\frac{1}{t}  )^{\lambda}\rbrace\\  
=&\dfrac{\Gamma(\frac{1}{\alpha}+1)}{\Gamma(\frac{1}{\alpha}+1-\xi)}&(1+bz^{\alpha}   )^{\lambda-\xi}\left(\frac{1}{t}\right)^{\frac{1}{\alpha}-\xi},
\end{eqnarray}
where $b'= bx^{\alpha}$.  By equating  Eq.~(\ref{ufo3}) and Eq.~(\ref{ufo4}),  one find that they match each other, yielding to:  
\begin{equation*}
\begin{split}
&\xi=1+\lambda+1/\alpha,\nu=-1-\frac{2}{\lambda+1/\alpha},\\
&\gamma=(1/\alpha-\lambda)(1+\frac{1}{\lambda+1/\alpha}).
\end{split}
\end{equation*}
Therefore, we see that the solution is:
\begin{equation}\label{ufo12}
P(x,t)=A^{NL}_{q}t^{\frac{-1}{\alpha}}\left[1+(1-q) \eta ^{NL}_{q} (\alpha,\lambda,D)\frac{x^{\alpha}}{t}\right]^{\lambda},
\end{equation}
where the pre-factor $A^{NL}_{q}$ is a normalization constant, and $\eta ^{NL}_{q} (\alpha,\lambda,D)$ is a constant depending on $A^{NL}_{q}$. If we again suppose that the distribution is symmetric with respect to $x$. Therefore, $x\rightarrow |x|$ (i.e. its absolute value), then the normalization constant is:
\begin{equation*}
A^{NL}_{q}= \frac{1}{2}\eta^{\frac{1}{\alpha}}(\alpha,\lambda,D) \dfrac{\Gamma(-\lambda)}{\Gamma(-\lambda-\frac{1}{\alpha})\Gamma(1+\frac{1}{\alpha})}, 
\end{equation*}
where $\eta  (\alpha,\lambda,D)=\dfrac{ \eta^{NL}_{q}(\alpha,\lambda,D)}{\lambda}$. From this expression one can calculate the final expression of $\eta^{NL}_{q} (\alpha,\lambda,D)$

\begin{equation}
\begin{aligned}
&\eta^{NL}_{q}(\alpha,\lambda,D)=...\\...&\lambda \left[ 2^{1-\nu}\alpha^{\nu}D B^{\nu} \left(-\lambda-\frac{1}{\alpha},\frac{1}{\alpha}\right)\Gamma(\lambda\nu+1)\Gamma \left(-\lambda-\frac{1}{\alpha}\right)\right]^{1/\xi},
\label{eq:nq}
\end{aligned}
\end{equation}
 where $ B(.,.) $ is the Beta function. By defining $ \lambda=\frac{1}{1-q}$ in Eq.~(\ref{ufo12}) (where $q>1$), we recover the \textit{first generalized} $q$-Gaussian distribution with two free independent parameters. We named Eq.~(\ref{ufo12}) as the non- local $q$-Gaussian (NL$q$-Gaussian) distribution.  \\
 The NL$q$-Gaussian distribution Eq.(\ref{ufo12}) is the Green function of Eq.(\ref{ufo1}) obtained using the fractional derivatives of  Riemann Liouville and Caputo, (non-local fractional definitions), for time and space, respectively. The NL$q$-Gaussian has been defined as  $g_{q}^{\lambda,\alpha}(x)$, Equation N.$8$ in Table \ref{table:1}. 
 The plots for NL$q$-Gaussian solutions are shown in Figure~\ref{fig:NLqGaussian}, for various $\alpha$ and $q$ values. With respect to the local case, the behavior for the non-local case is more complicated. As is seen from the subfigure \ref{fig:NLqGaussian}a, the case where $\alpha$ is kept constant and $q$ increases is evaluated. For $q<1.3$, the peak rises, and for $q\geq 1.3$, it decreases. In subfigure \ref{fig:NLqGaussian}b, for a constant $q$, however, the peak increases when $\alpha$ increases. The subfigure \ref{fig:NLqGaussian}c shows the time evolution of the Green function of Eq.~\ref{ufo1}, which solution is the NL$q$-Gaussian distribution. The example was made for the values of $q=1.5$ and $\alpha=2 $. We have shown that the distribution widens as time goes on. 
 
%%%%%%%%%%%%%%%%%%%%%%%%%%%%%%%%
\section{AN Application of L$q$-Gaussian in S\& P500 Stock Markets }\label{SEC:application}
The price return in stock markets exhibits remarkable characteristic features. The most largely observed feature in recent studies is the  self similarity law , where the  PDF  obeys:
\begin{equation}
P(x,t)=\frac{1}{(B t)^{H}} f \left( \dfrac{x}{(B t) ^{H}} \right),  
\label{eq:P-Gaussian}
\end{equation}

\begin{table*}[ht]
 \begin{tabular}{| P{2.2cm}| P{3.5cm}|P{3.9cm} |P{3.4cm}||P{4.3cm}|} 
 \hline
 & \makecell{Classical \\ (C-PME)}  & \makecell{Time fractional \\ (T-FPME) }  & \makecell{Time-Space fractional \\(TS-FPME)} & Particular case of the Generalized PME (G-PME )\\
 \hline
Forms of PME & \begin{equation} \nonumber \begin{aligned}   \dfrac{\partial P}{\partial t}=&D\dfrac{\partial P^{2-q}}{\partial x^{2}}, \\ 1<&q<3 \end{aligned}  \end{equation} &  \begin{equation} \begin {aligned}\dfrac{\partial^{\xi} P(x,t)}{\partial t^{\xi}}=D&\dfrac{\partial^{2} P (x,t)^{2-q}}{\partial x^{2}}, \\
0<\xi& \leq 1,\\  1<q& < 3
\end{aligned} \label{Eq:Self_similar}\end{equation} & \begin{equation} \begin{aligned}\dfrac{\partial^{\xi} P}{\partial t^{\xi}}=D&\dfrac{\partial^{\gamma} P^{2-q}}{\partial x^{\gamma}},\\
0<\xi &\leq 1, \\ 1<\gamma &\leq 2,\\1<q& < 3
\end{aligned} \nonumber \end{equation}& \begin{equation}\nonumber
\begin{aligned}_{0}\mathcal{D} ^{\xi,\frac{1}{t}}P= D & \left(^{C} \mathcal{D}_{1}^{\gamma, x^{\alpha}}P^{\nu}\right),\\ \xi,\gamma&>0\end{aligned}\end{equation}  \\ 
 \cline{1-4}
Parameters after comparing with General PME &$\begin{aligned}
 q &=2-\nu,\\
 \gamma&=2,\\
 \xi&=1
 \end{aligned}$ & $\begin{aligned}
  q &=2-\nu,\\
 \gamma&=2,\\
 \xi &=\dfrac{3-q}{\alpha}
 \end{aligned}$ & $\begin{aligned}
 q &=2-\nu,\\
 \gamma&>1,\\
 \xi&=\dfrac{1-q+\gamma}{\alpha}
 \end{aligned}$ & $\begin{aligned}
\xi&=1+\lambda+1/\alpha, \\ \nu&=-1-\frac{2}{\lambda+1/\alpha},\\
\gamma&=(1/\alpha-\lambda)\left(1+\frac{1}{\lambda+1/\alpha} \right)
 \end{aligned}$ \\ 
 \hline
\end{tabular}
\caption{Summary of particular forms of the generalized PME Eq.~\ref{Eq:Matrix} to model the time evolution of the PDFs of price return. The C-PME, T-FPME, TS-FPME and G-PME are obtained after set a specific value of the parameters in the general form of the PME. The fittings were perform by setting the parameters as shown in the table. }
\label{table:2}
\end{table*}
 in which  $f$ is a normalized distribution that is usually fit to a $q$-Gaussian~\cite{ref8}. In earlier work, the function $f$ was assumed as a Levy-stable distribution function, $L_{\gamma}$, which has the drawback that it presents infinite standard deviation and it does not obey the empirical power law tails \cite{Gopi1998,Repe2004}. 
\begin{figure}[!htbp]
\centering
\includegraphics[scale=0.37,trim=-0.5cm 0cm 2cm 1cm]{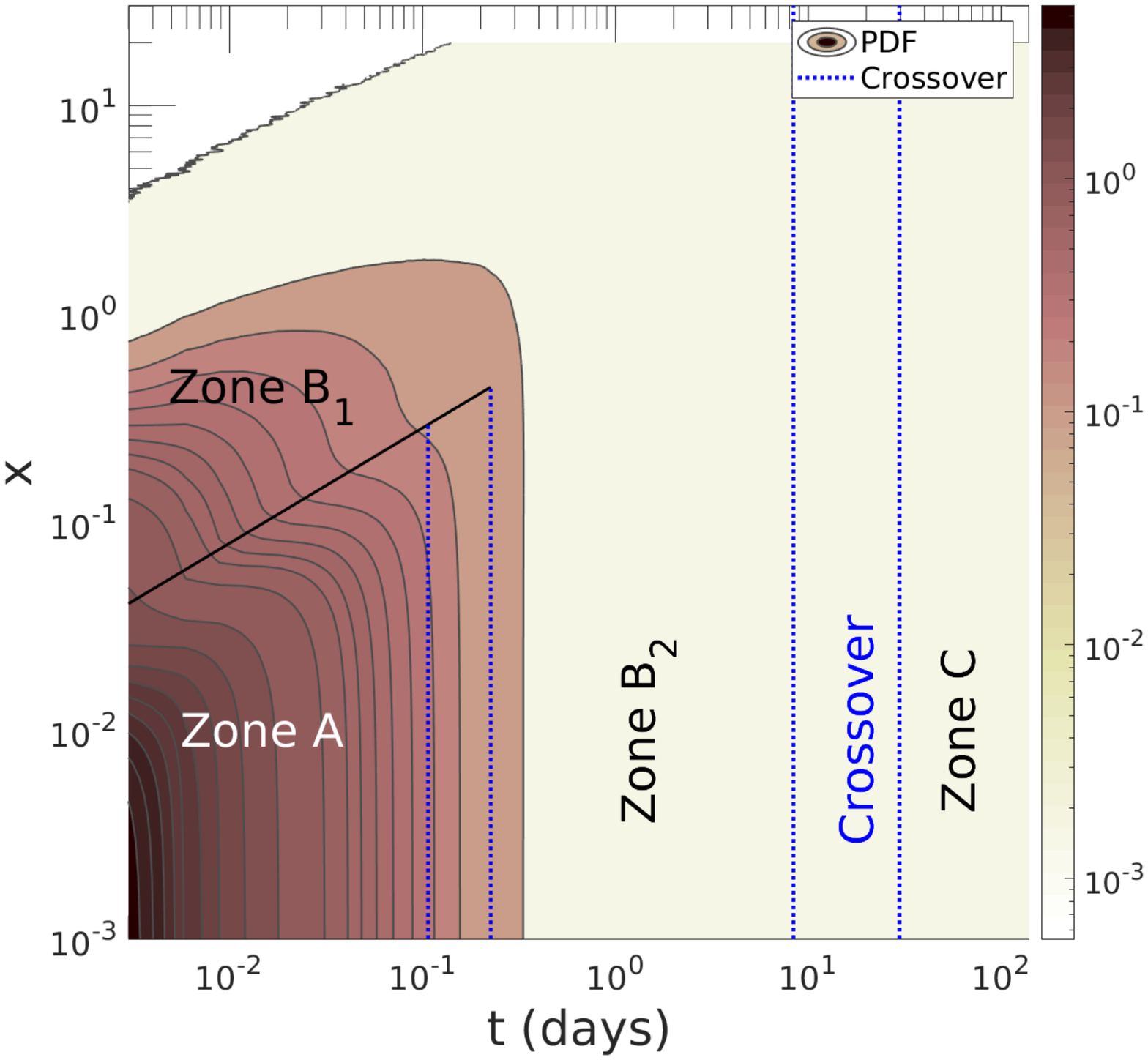}
\caption{Three different zones were determined based on an
abrupt slope change of the fitting parameters $\alpha$ and $q$. The contour plot represents the PDF of the detrended price return. The black circles represent the end points of the strong super-diffusion regime (zone A)
from $t=0$ to $t=35$ minutes. These points are the ends of the bump obtained from the two points at the PDF with an abrupt change of slope (Figure 1-c in~\cite{Alonso}).
The remaining area during the first $35$ minutes corresponds to a weak super-diffusion regime (Zone $B_{1}$). From $80$ min to $10$ days, the zone corresponds to a weak super-diffusion
regime (zone $B_{2}$). A normal diffusion process is reached after
$30$ days. The gray dashed lines represent the transitions
between each zone.}
\label{fig:Regimes}
\end{figure}
For long time returns it will be proved that $f$ is a Gaussian distribution function, where the price return behaves like independent and identically distributed random variables but still following the self similar principle. Particular cases of the generalized PME  are presented in Table \ref{table:2} . The solutions of each of these partial differential equations obey the self-similar law presented in Eq.(\ref{eq:P-Gaussian}) and are related to the q-Gaussian distribution function.\\

In this part, we analyze  the S\&P500 stock market data during the 24-year period from January 1996 to August 2020  with a frequency of one minute. The detrended price return is defined as,
\begin{equation}
x(t)=I^{*}(t_{o}+t)-I^{*}(t_{o}),
\label{eq:Index}
\end{equation}
where $I^{*}(t_o)$ is the detrended stock market index at time $t_o$, and $I^{*}(t)$ is the detrended stock market index for any time  $t>t_o$. The PDF  of the detrended price return has been fitted using the q-Gaussian distribution \cite{ref8}, where different zones had been captured from strong to super diffusion regime previously \cite{ref8,ref20}.  Figure \ref{fig:Regimes} shows the evolution of the PDF and its regimes in the $(x,t)$ space. Initially, the PDF has a pronounced bump in the center that fully disappears close to $80\,min$. During the first $35\,min$ there is a power law relation of time against the end points of the bumps at the PDFs (See Subfigure 1-e in~\cite{Alonso}).  This zone is defined as the strong super diffusion regime (Zone A). The remaining area during that time and the following next area close to $10$ days corresponds to the weak super diffusion regime (zone B). Finally, the last regime corresponds to a normal diffusion process (Zone C) and is reached after $30$ days approximately.

\begin{figure*}[!htbp]
\centering 
\includegraphics[scale=0.40,trim=0cm 0cm 0cm 15cm]{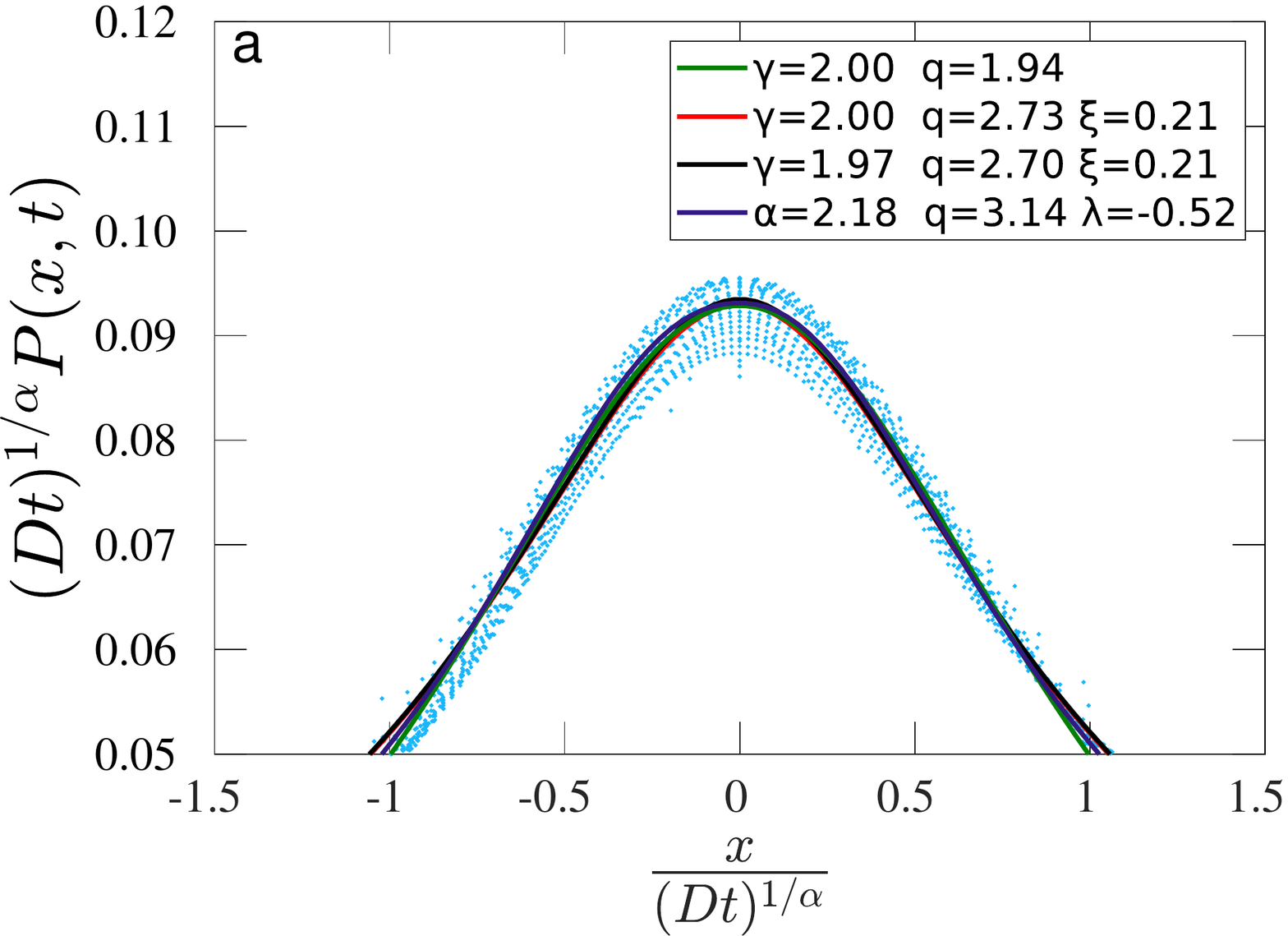}
\includegraphics[scale=0.40,trim=0cm 0cm 0cm 15cm]{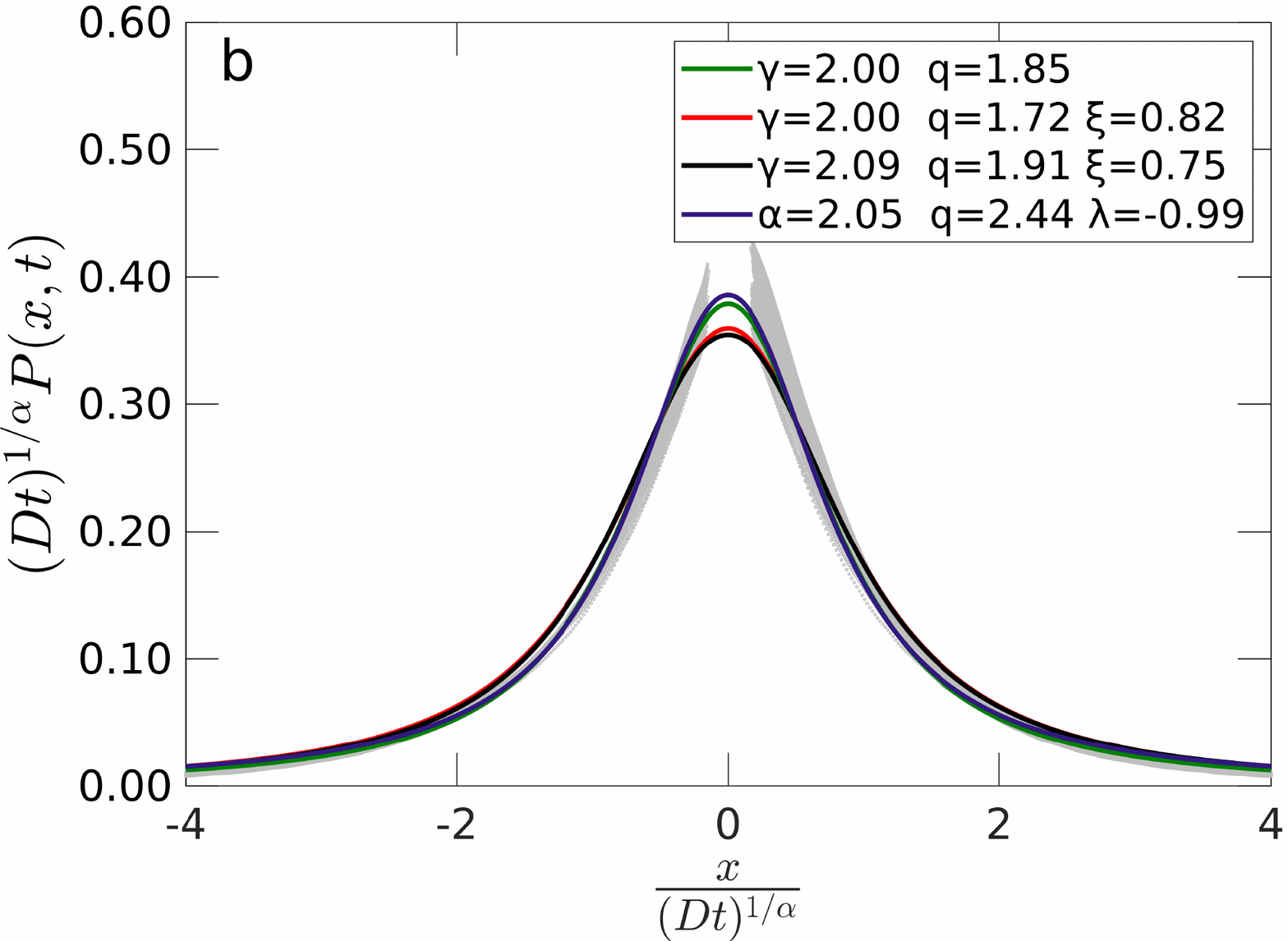}
\includegraphics[scale=0.40,trim=0cm 0cm 0cm 15cm]{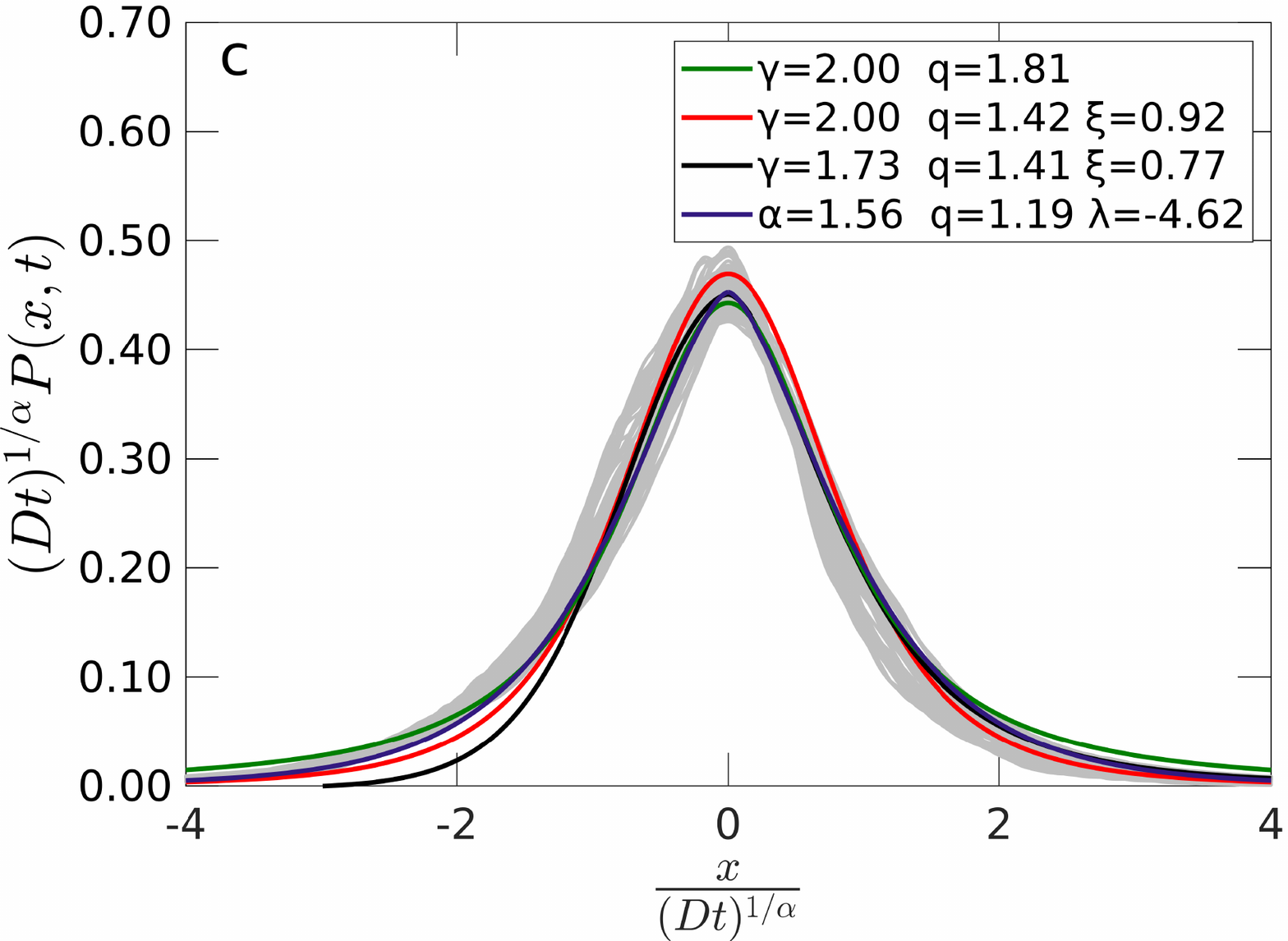}
\includegraphics[scale=0.40,trim=0cm 0cm 0cm 15cm]{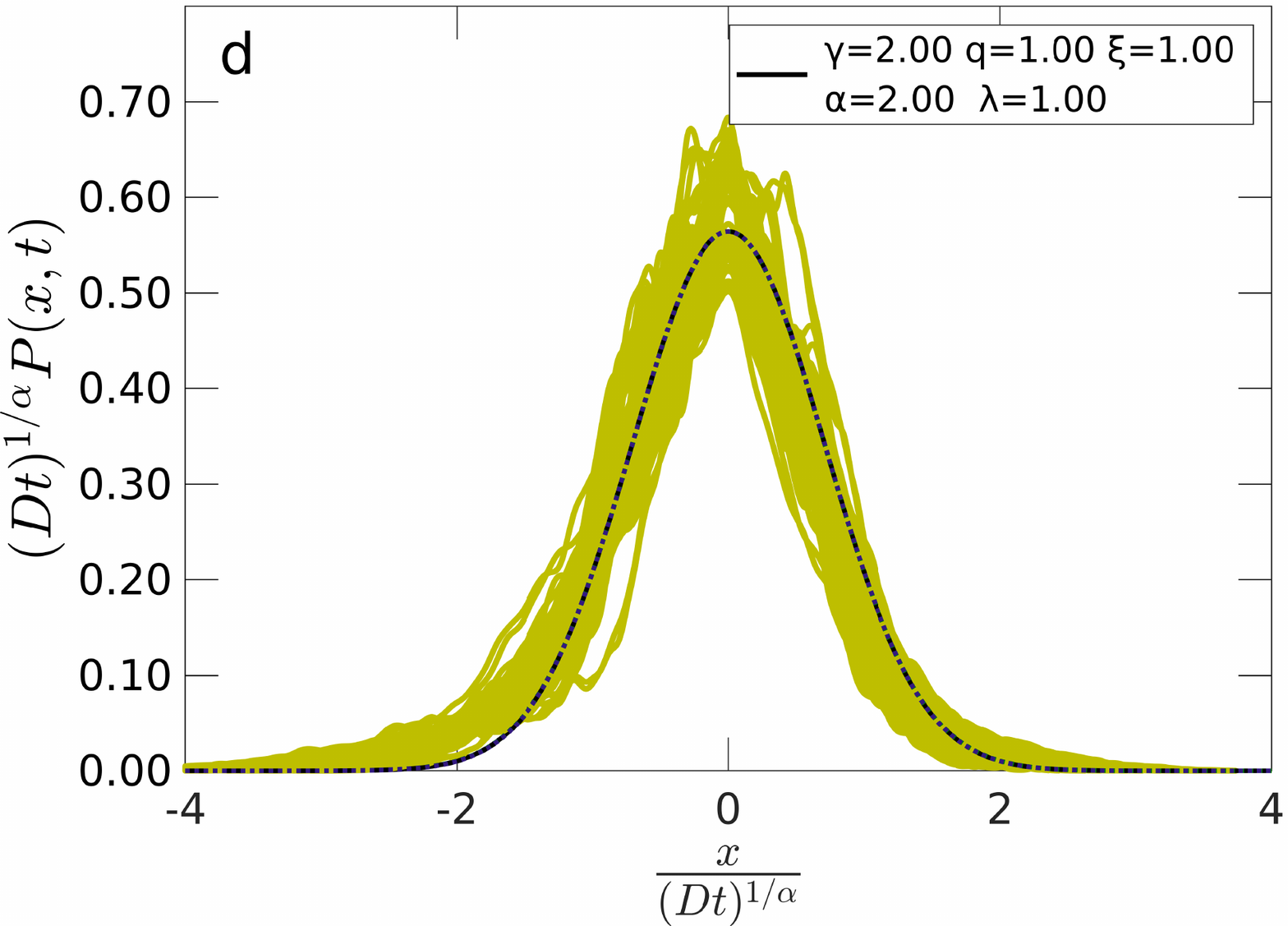}
\caption{This figure shows the collapse of the time evolution of the PDFs for each zone displayed previously in Figure \ref{fig:Regimes}. (a) The collapse of the PDFs of the strong superdiffusion regime (Zone A), (b) The collapse of the PDFs of the weak superdiffusion regime for the first $35$ $min$ while the bump remains in the PDFs of Zone $B_{1}$. 
%A first example of the analytic continuation is observed here in which $\gamma>2$ is out of the range that we considered at the beginning. 
(c) The collapse of the PDF's during the weak superdiffusion regime after $100$ $min$, when the bump disappears completely in Zone $B_{2}$, and (d) The collapse of the PDFs for the normal diffusion regime (Zone C). Each of the collapsed data is fitted by the solution of four different $q$-Gaussian forms, which are the solutions for the partial differential equations provided in Table I: no-fractional(green), time-fractional(red), time-space fractional (black), and a particular form of the G-PME (purple). For the subfigure (d) the collapsed data is fitted by a Gaussian distribution function, which is a concurrent solution for the four previous PDEs when $q=1$ and $\gamma=1$.}
\label{fig:SPStockMarkets}
\end{figure*}

We have reconstructed the  time  evolution of  the PDFs of the detrended price return and we collapsed them after  applying  the corresponding re-scaling factor. Four equations of the Table \ref{table:1} have been used to model this behaviour. The Classical PME (C-PME), the time  FPME (T-FPME), the time-space   FPME (TS-FPME) and a particular case of the generalized PME (G-PME). These solutions obey Eq.~(\ref{eq:P-Gaussian}), and are presented in Table \ref{table:2} with more detail.

  The TS-FPME and the particular solution of the G-PME proposed in this manuscript are new options to model the time evolution of the detrended price return. The L$q$-Gaussian  and NL$q$-Gaussian, which are the solutions of the TS-FPME and the particular case of the G-PME respectively, fit the collapse of the PDFs of the detrended price return well. Figure \ref{fig:SPStockMarkets} shows the result of these fittings.  

The first best option to fit the price return was obtained by replacing $f$ as the L$q$-Gaussian ($g_{q}^{\gamma}$) in Eq.~(\ref{eq:P-Gaussian}), this equation can be written as:
\begin{equation}
P(x,t)=\dfrac{1}{(Bt)^{\frac{\xi}{1-q+\gamma}}} \dfrac{1}{C_{q}^{\gamma}} {\left(1- (1-q) \dfrac{x^{\gamma}}{(Bt)^{\frac{\gamma \xi}{1-q+\gamma}}} \right)^{\dfrac{1}{1-q}}},
\label{L-q_Gaussian}
\end{equation}
where $C_{q}^{\gamma}$ is the normalization constant. $B$ is related with the diffusion term, both of them are detailed in Table \ref{table:1}.

The second best option was obtained by replacing $f$ as the NL$q$-Gaussian ($g_{q}^{\alpha,\lambda}$) in Eq.~(\ref{eq:P-Gaussian}), this second equation can be written as:
\begin{equation}
P(x,t)=\dfrac{1}{(Bt)^{\frac{1}{\alpha}}} \dfrac{1}{C_{q}^{\lambda,\alpha}} {\left(1- (1-q) \dfrac{x^{\alpha}}{(Bt)} \right)^{{\lambda}}}.
\label{NL-q_Gaussian}
\end{equation}

%The TS-PME is a generalized form of the Porous Media Equation, and its solution is the L$q$-Gaussian, which is the \textit{first generalized} form of the $q$-Gaussian function. The L$q$-Gaussian, Eq.(\ref{eq:p_x}),  can be expressed as:

%The Eq.~(\ref{Eq:Matrix}) is the G-PME and it is a more general form of the PME equation with the form:
%\begin{equation} \nonumber
%_{a}\mathcal{D}_{t}^{\xi, t^{n}}P(x,t)= D  \left(^{C}_{x}\mathcal{D}_{b}^{\gamma, x^{\alpha}}P^{\nu}(x,t)\right),
%\end{equation}
%The solution is provided when $n=-1$, $a=0$, and $b=1$ in Eq.(\ref{ufo12}) and it is called NL$q$-Gaussian. The NL$q$-Gaussian can be expressed as:

%where
%\begin {equation}
%\begin{aligned}\
 %\xi&=1+\lambda+1/\alpha, \\ %\nu&=-1-\frac{2}{\lambda+1/\alpha},\\
%\gamma&=(1/\alpha-\lambda) %\left(1+\frac{1}{\lambda+1/\alpha}\right)
%\end{aligned}
%\end{equation}
The $C_{q}^{\lambda,\alpha}$ is the normalization constant, and $B$ is related with the diffusion term. The definitions of these parameters are expressed in Table \ref{table:1}.
By considering $\lambda=\dfrac{1}{1-q}$, the Eq.~(\ref{L-q_Gaussian}) is recovered. \par

 The results of fitting the collapse of the PDF of the detrended price return are shown in Figure~\ref{fig:SPStockMarkets}. The Figure \ref{fig:SPStockMarkets} presents the collapses of the PDFs of price return for the specific zones presented in Figure \ref{fig:Regimes}. Each collapse has been fitted by the four solutions of the equations presented in Table \ref{table:2}. 
 The best fitting for the four cases is the NL$q$-Gaussian. However, the four solutions constitutes an acceptable solution for the correspondent collapses of the PDFs. A converge to a Gaussian normal distribution is observed for long time returns.

%%%%%%%%%%%%%%%%%%%%%%%%%%%%%%%%

\section{Conclusions}

We  provided  different solutions for the generalized form of the FPME. To this end, we had considered the generalized PME (G-PME) as the master equation. We  introduced  the anomalous PME as a nonlinear fractional diffusion equation, which is a particular case of G-PME. The solutions were built by considering the local and non-local fractional derivatives assuming a Dirac’s delta function as the initial condition. Our analysis proved that the solution are given by a generalized $q$-Gaussian, which obey a self similar law. \\
The fractional derivatives were classified as local and non-local, where the Katugampola's is the local fractional derivative and the Riemman-Lioville, Caputo and Riesz are non-local fractional derivatives.\\
First, we  considered  the case where the derivatives are local. The resulting solution is L$q$-Gaussian, which is the \textit{first generalized} $q$-Gaussian function.  This solution fits the PDF of the detrended price return well  (Sec.~\ref{SEC:application}). The second analyzed class of G-PME is the one in which the time and space derivatives are given by  the  non-local fractional generalizations: Riemann-Lioville  and Caputo, both of them based on the Laplace transform. For this second class, the fractional derivatives are evaluated  with respect to another function, and proved that they admit the \textit{second generalized} $q$-Gaussian solution. This second solution is symmetric about its mean (peak). The NL$q$-Gaussians hold a different self-similar law   than  the L$q$-Gaussians. The main difference is that the self-similarity of the L$q$-Gaussians depends on $\alpha$ and $\gamma$ only, for the NL$q$-Gaussians the self-similarity depends on $\alpha$, $\gamma$ and one extra global exponent $\lambda$, where the \textit{first generalized} $q$-Gaussian is recover for $\lambda=\frac{1}{1-q}$.
 A  hybrid equation has also been considered, where the time  dimension  is local fractional derivative, and the  spatial dimension  is non-local. The solution that we found is again proportional to the generalized $q$-Gaussian (which we named LNL$q$-Gaussian), but they obey a power in $x$ that causes the PDF to vanish in the limit $x\rightarrow 0$. The shape of these PDFs are very different from the $q$-Gaussian distribution and  they are not symmetrical distribution.\\

%The generalized $q$-Gaussian functions found are a class of analytic solutions of (local and non-local) fractional PMEs, these solutions will be useful in other problems involving non-linear anomalous diffusion.

%(KARINA, THE TEXT BELOW IS BROKEN)
 
The L$q$-Gaussian (\textit{first generalized} q-Gaussian) and NL$q$-Gaussian (\textit{second generalized} q-Gaussian) have been used to model the detrended price return of S\&P500. Both distribution functions describe well the fitting of the detrended price return. The NL$q$-Gaussian is the best model to fit the probability of the detrended price return. For future work the generalized form of the PME will be solved by applying the $q$-Fourier analysis. The ordinary Fourier analysis only applies for linear operators. The generalized PME contains nonlinear operators, preventing us from using the ordinary Fourier analysis. 

%Further research is required to find the solution of the time-space dependent fractional PME.

%for the following reason: Let $\hat{O}_{\text{L}}$ be a linear differential equation. Then if $Y_1$ and $Y_2$ are two independent solutions, then any linear combination of $Y_1$ and $Y_2$ are also solutions, i.e. if $\hat{O}_{\text{L}}(Y_1)=0$, and $\hat{O}_{\text{L}}(Y_2)=0$, then $\hat{O}_{\text{L}}(aY_1+bY_2)=0$ where $a$ and $b$ are two real complex numbers. For nonlinear operators however this is not the case, i.e. if $\hat{O}_{\text{NL}}$ is a non-linear operator, and $\hat{O}_{\text{NL}}'(Y1)=0$, and $\hat{O}_{\text{NL}}'(Y2)=0$, then one cannot conclude $\hat{O}_{\text{NL}}'(aY1+bY2)=0$. This prevents us from using the ordinary Fourrier analysis for solving the Eq.(\ref{general eq.}). %for which one solves the problem for a e.g. plane wave as the Fourrier components, and then sum them up together to have the solution in the direct space.
%To be more precise, let $e^{ikx}$ is the solution in the Fourier space. Then there is no guarantee that $\sum_k a(k)e^{ikx}$ is also a solution, because of the nonlinear nature of the governing equation. This problem will be resolved for the $q$-Fourier analysis, for which the linearity is translated in a different way.
%%%%%%%%%%%%%%%%%%%%%%%%%%%%%%%%
\appendix

\section{Properties of Katugampola derivative}\label{SEC:Katugampola-derivative}

In here,  we give a brief summary of the definition of the Katugampola fractional operator  and some of its properties.
This local fractional operator  is used to construct the TS-FPME in sec.~\ref{SEC:local-local}.
If $0 \leq \alpha<1$, the Katugampola operator generalizes the
classical calculus properties of polynomials \cite{Katugampola2014}. Furthermore, if $\alpha = 1$, the definition is equivalent to the
classical definition of the first order derivative of the function $f$. 
 The Katugampola derivative is defined as:
\begin{equation}
\mathcal{D} ^{\alpha}f(t)=\lim_{\epsilon \to 0}\dfrac{f(te^{\epsilon t^{-\alpha}})-f(t)}{\epsilon}
\label{Eq:Katugampola}
\end{equation}
for $ t>0  $ and $ \alpha\in(0,1]$. 
When $\alpha \in (n, n+1]$ (for some $n \in \mathbb{N}$, and $f$ is an $n-$differentiable at $ t>0,$), the above definition generalizes to
\begin{equation*}\label{ufo8}
\mathcal{D} ^{\alpha}f(x)=\lim_{\epsilon \to 0}\dfrac{f^{(n)}(x e^{\epsilon x^{n-\alpha}})-f^{(n)}(x)}{\epsilon}, 
\end{equation*}
and if $f$ is $(n+1)-$differentiable at $t>0, $ then
\begin{equation*}
\mathcal{D} ^{\alpha}f(t)=t^{n+1-\alpha}f^{(n+1)}(t).
\end{equation*}
In continue, we review some properties of the Katugampola derivative in Table \ref{table:4}. If $ f $ is $\alpha-$ differentiable in some $(0, a),\, a > 0,$ and $f^{(\alpha)}(0)= \lim_{t \to 0^{+}}\mathcal{D} ^{\alpha}f(t)$ exists, the following properties hold for Katugampola derivative. For $ f, g $, be $\alpha-$ differentiable at a point $ t > 0.$  
%Let $f$ be a function with the Katugampola derivative. By using the recent properties in Table \ref{table:4} of the Katugampola derivative, we can write
\begin{align*}
\mathcal{D} ^{\alpha}f(at) =& f^{\prime}(at)\mathcal{D} ^{\alpha}(at)\\
&= af^{\prime}(at)\mathcal{D} ^{\alpha}t\\
&= af^{\prime}(at)t^{1-\alpha}\\
&=  (at) ^{1-\alpha}a^{\alpha}f^{\prime}(at).
\end{align*}
One can define the inverse of the $\mathcal{D} ^{\alpha} $ operator as a fractional integral,
\begin{equation*}
( \mathcal{D}^{\alpha})^{-1} \equiv \mathcal{D}^{-\alpha}\equiv I^{ \alpha}=\int  ^{t} dx\frac{(.)}{x^{1-\alpha}},
\end{equation*}
where the $(.)$ symbol is serving as place holder for the function to be operated upon. One verifies that
\begin{equation*}
\mathcal{I}^{\alpha}[\mathcal{D}^{ \alpha}( f )]=\int  ^{t} dx\frac{x^{1-\alpha}f^ {\prime} }{x^{1-\alpha}}=f,
\end{equation*}

where $f$  vanishes at the lower limit. Then:
\begin{align*}
\mathcal{D}^{\alpha}[\mathcal{I}^{ \alpha}( f )]&= \mathcal{D}^{\alpha}[\int  ^{t} dx\frac{ f }{x^{1-\alpha}}]\\
&=t^{1-\alpha} \left(\int  ^{t} dt\frac{ f }{x^{1-\alpha}}\right)^{\prime}\\
&=t^{1-\alpha}\frac{ f }{t^{1-\alpha}}=f.
\end{align*}

\section{Caputo fractional derivative of a function with respect to another function}\label{SEC:Caputo-derivative}

This section contains definitions  of non-local fractional operators that are used in this paper to construct  the TS-FPMEs. The  
Riemann-Liouville  fractional derivative is a fractional operator that is used in Sections~\ref{SEC:L-N} and ~\ref{SEC:N-N} as a  non-local fractional operator to construct the TS-FPMEs   with  (LN) and (NN) fractionalizations. The integral representation for this operator is:
\begin{equation*}
^{RL}_{a}\mathcal{D} ^{\alpha,x}f(t)=\frac{1}{\Gamma(n-\alpha)}\frac{d^{n}}{dx^{n}}\int_{a}^{x}\dfrac{dtf(t)}{(x-t)^{\alpha+1}}, 
\end{equation*}
where $n-1<\alpha \leq n$ ~\cite{ref21}. A recent variation of RL operator is the Caputo derivative~\cite{Caputo}, defined as:
\begin{equation*}
^{C}\mathcal{D}_{b} ^{\alpha,x}f(t)=\frac{(-1)^{n}}{\Gamma(n-\alpha)} \int_{x}^{b}\dfrac{dtf^{(n)}(t)}{(x-t)^{\alpha+1-n}},\,\,\,\,\,n-1<\alpha \leq n,
\end{equation*}
where $C$ stands for Caputo and $f^{(n)}$ is the $n$th derivative of $f$. The main advantage of the Caputo derivative is that the derivate of a constant is zero, which is not the case of the RL operator. Substantially, this kind of fractional derivative is a formal generalization of the integer derivative under Laplace transform~\cite{Bologna}.\par A generalized fractional operator  that we used to construct the TS-FPME with (NN) fractionalization  in Sec.~\ref{SEC:N-N} is the Caputo fractional derivative of a function with respect to another function~\cite{ref19}, and defined as:
\begin{align*}
 ^{C}&\mathcal{D}_{b} ^{\alpha,\psi(x)}f(t)=\frac{(-1)^{n}}{\Gamma(n-\alpha)}...\\
&...\times\int_{x}^{b}\psi(t)^{\prime}(\psi(t)-\psi(x))^{n-\alpha-1 }(\frac{1}{\psi(t)^{\prime}} \frac{d}{dt} )^{n} f(t)dt.
\end{align*}
Note that  the recent integral representation  in  the special case  $\psi(x)=x $  is reduced  to the integral representation of the Caputo derivative.
\\
We solve a particular case of the generalized PME, Eq.~(\ref{Eq:Matrix}), described by a fractional derivative of a function with respect to another function.  This innovative approach will be useful to solve other physical problems that present a self-similar pattern and can be modelled by a $q$-Gaussian. 
\section{Fractional derivatives: Definition and properties }\label{SEC:Definition-Properties}
In this section, we give a short review of the properties of the fractional derivatives: Katugampola, Riemann-Lioville, and Caputo. The Katugampola is one definition for the local fractional derivative. The Riemann-Lioville and Caputo are definitions of non-local derivatives. A comparison between each property  of these fractional derivatives are presented in Table \ref{table:4} .\\
Katugampola's corresponds to the ordinary derivative when $\alpha=0$ and $\alpha=1$. The Riemann-Lioville and Caputo are an analytical continuation of the ordinary derivatives. The main difference between them is that the Caputo derivative of a constant is zero, a property that does not hold for Riemann-Lioville derivative. This desirable property is satisfied by the Katugampola local derivative, too.

\renewcommand{\arraystretch}{4}
\begin{table*}[ht]
\centering
 \begin{tabular}{| P{1.20cm}| P{5.1cm}| P{6.8cm}| P{5.15cm}| } 
 \hline
 Property & Katugampola \cite{Anderson2015,Katugampola2014}   & Riemann-Lioville \cite{Ishteva2005,Khader2015,He2012,ref19} & Caputo \cite{Ishteva2005,Khader2015,He2012}\\
  \hline
Key Property & $\mathcal{D} ^{\alpha}f( t)=\lim_{\epsilon \to 0}\dfrac{f(te^{\epsilon t^{-\alpha}})-f(t)}{\epsilon}$  & $^{RL}{\mathcal{D}}^{\alpha}f(t)={\mathcal{D}}^{n}{\mathcal{I}}^{n-\alpha}f(t)$ & $^{C}{\mathcal{D}}^{\alpha}f(t)={\mathcal{I}}^{n-\alpha} {\mathcal{D}}^{n}f(t)$\\
\cline{3-4}
& $\mathcal{D} ^{\alpha}f(t) =t^{1-\alpha}\dfrac{df(t)}{d t}$ & \multicolumn{2}{c|}{$\quad \quad  n-1<\alpha<n, \quad\quad {\mathcal{D}}^{n}=\dfrac{d^{n}}{dt^{n}},\,\,\,\,  \quad\quad {\mathcal{I}}^{\alpha}f(t):=\dfrac{1}{\Gamma(\alpha)}\bigintss_{a}^{t}f(\tau)(t-\tau)^{\alpha-1}d\tau$\,\,\,\,\,}\\
  \hline
 Cte. function & $\mathcal{D} ^{\alpha}c=0$  & $^{RL}{\mathcal{D}}^{\alpha}c=\dfrac{c}{\Gamma(1-\alpha)}t^{-\alpha}$ & $^{C}{\mathcal{D}}^{\alpha}c=0$  \\

 \hline
 Linearity &\multicolumn{3}{c|}{ ${\mathcal{D}}^{\alpha}(a f(t)+g(t))=a{\mathcal{D}}^{\alpha}f(t)+\mathcal{D}^{\alpha}g(t)$} \\
 \hline
Product (Leibniz)& $\mathcal{D} ^{\alpha} (f(t)g(t))=f(t)\mathcal{D} ^{\alpha}g(t)
 +g(t)\mathcal{D} ^{\alpha}f(t)$& $^{RL}{\mathcal{D}}^{\alpha}(f(t)g(t))=\mathlarger{\sum}_{k=0}^{\infty} \binom{\alpha}{k}(^{RL}\mathcal{D}^{\alpha-k}f(t))g^{k}(t)$ & $\begin{aligned}^{C}{\mathcal{D}}^{\alpha}(f(t)g(t))=^{\,\,\,\,\,RL}{\mathcal{D}}^{\alpha}(f(t)g(t))....\\...-\mathlarger{\sum}_{k=0}^{n-1} \dfrac{t^{k-\alpha}}{\Gamma(k+1-\alpha)}((f(t)g(t))^{k}(0))\end{aligned}$  \\
 \hline
Quotient Rule& $\mathcal{D} ^{\alpha} \left(\dfrac{f(t)}{g(t)}\right)=\dfrac{g(t)\mathcal{D} ^{\alpha}f(t)-f(t) \mathcal{D} ^{\alpha}g(t)}{g(t)^{2}}$&  &  \\
 \hline
 Chain Rule& $\mathcal{D} ^{\alpha}\left(fog\right)   =\dfrac{df}{dg} \mathcal{D} ^{\alpha}g(t)$&  $\begin{aligned} &^{RL}_{\,\,\,a}{\mathcal{D}} ^{\alpha,x}(fog)=\\ &\dfrac{1}{\Gamma(1-\alpha)} \left( \dfrac{1}{g'(x)} \dfrac{d}{dx}\right)^{n} \bigintss_{a}^{x}\dfrac{g'(\tau) f(\tau) d\tau }{\left[ g(x)-g(\tau)\right]^{1+\alpha-n}}\\
 &^{RL} {\mathcal{D}}_{b}^{\alpha,x}(fog)=\\ &\dfrac{(-1)^{n}}{\Gamma(1-\alpha)} \left( \dfrac{1}{g'(x)} \dfrac{d}{dx}\right)^{n} \bigintss_{x}^{b}\dfrac{g'(\tau) f(\tau) d\tau }{\left[ g(\tau)-g(x)\right]^{1+\alpha-n}}
 \end{aligned}$ & $\begin{aligned} &^{C}_{\,\,\,a}{\mathcal{D}} ^{\alpha,x}(fog)=\\ &\dfrac{1}{\Gamma(1-\alpha)}  \bigintss_{a}^{x}\dfrac{g'(\tau) f^{(n)}_{g}(\tau) d\tau }{\left[ g(x)-g(\tau)\right]^{1+\alpha-n}},\\
 &^{C} {\mathcal{D}}_{b}^{\alpha,x}(fog)=\\ &\dfrac{(-1)^{n}}{\Gamma(1-\alpha)} \bigintss_{x}^{b}\dfrac{g'(\tau) f^{(n)}_{g}(\tau) d\tau }{\left[ g(\tau)-g(x)\right]^{1+\alpha-n}},\\
 &f_{g}^{n}(\tau)=\left( \dfrac{1}{g'(x)}\dfrac{d}{dx}\right)^{n}f(\tau)
 \end{aligned}$  \\
 \hline
 Power function& $\mathcal{D} ^{\alpha}(t^{p})=p t^{p-\alpha}$& $\begin{aligned}^{\,\,\,RL}{\mathcal{D}}^{\alpha}(t^p)&=\dfrac{\Gamma(p+1)}{\Gamma(p-\alpha+1)}t^{p-\alpha},\\ &p>-1,p\in \mathbb{R}\end{aligned}$ & $\begin{aligned}
^{C}{\mathcal{D}}^{\alpha}(t^p)&=^{RL}{\mathcal{D}}^{\alpha}(t^{p}), \\  p >& n-1,  p \in \mathbb{R}. \\ 
^{C}{\mathcal{D}}^{\alpha}(t^p)&=0, \\  p\leqslant& n-1, p \in \mathbb{N}
\end{aligned}$  \\
  \hline
  \end{tabular}
\caption{Comparison of properties between Katugampola, Riemman-Lioville and Caputo fractional derivatives. }
\label{table:4}
\end{table*}

%%%%%%%%%%%%%%%%%%%%%%%%%%%%%%%%
\vspace*{10cm}

\end{document}